\documentclass[preprint,12pt]{imsart}

\usepackage{amssymb,amsmath, subcaption}

\RequirePackage[OT1]{fontenc}
\RequirePackage{amsthm,amsmath}
\RequirePackage[numbers]{natbib}
\RequirePackage[colorlinks,citecolor=blue,urlcolor=blue]{hyperref}

\usepackage{amscd,amsfonts,amssymb,amsmath,latexsym,array,hhline,xcolor,graphicx}
\usepackage{float}
\def \a{\alpha}
\def \b{\beta}

\def \l{\lambda}

\def \p{\pi}

\def \s{\sigma}

\def \o{\omega}
\def \O{\Omega}
\def \e{\varepsilon}

\newcommand\F{\mbox{I\kern-2pt F}}

\newcommand\cF{{\cal F}}
\newcommand\cG{{\cal G}}

\newcommand\cB{{\cal B}}
\newcommand\cH{{\cal H}}
\newcommand\cK{{\cal K}}

\newcommand\cR{{\cal R}}

\newcommand\cP{{\cal P}}

\newcommand\R{{\mathbb{R}}}

\newcommand\E{\mathbb{E}}

\newtheorem{theo}{Theorem}[section]
\newtheorem{prop}[theo]{Proposition}
\newtheorem{lemm}[theo]{Lemma}
\newtheorem{defi}[theo]{Definition}
\newtheorem{ex}[theo]{Example}

\newtheorem{coro}[theo]{Corollary}
\newtheorem{rem}[theo]{Remark}
\newcommand\fdem{$\Box$}
\newcommand\beq{\begin{equation}}
\newcommand\eeq{\end{equation}}
\newcommand\bea{\begin{eqnarray}}
\newcommand\eea{\end{eqnarray}}
\newcommand\bean{\begin{eqnarray*}}
\newcommand\eean{\end{eqnarray*}}

\newcommand\bP{{\bf {\rm P}}}

\newcommand{\esssup}[1][\cH]{\mathrm{ess\,sup}_{#1}}
\newcommand{\essinf}[1][\cH]{\mathrm{ess\,inf}_{#1}}
\newcommand{\supp}[1][\cH]{\mathrm{ supp\,}_{#1}}

\usepackage[margin=3cm]{geometry}

\begin{document}
\begin{frontmatter}
\title{Pricing without no-arbitrage condition in discrete time}

\author[A1]{Laurence Carassus,} 
\author[A2]{Emmanuel L\'epinette,}

\address[A1]{ L\'{e}onard de Vinci P\^ole Universitaire, Research Center, 92 916 Paris La D\'{e}fense, France \\ and Laboratoire de Math\'{e}matiques de Reims, UMR9008 CNRS et Universit\'{e} de Reims Champagne-Ardenne, France \\ 
Email: laurence.carassus@devinci.fr}

 \address[A2]{ Paris Dauphine university, PSL research university, Ceremade,  CNRS, UMR, Place du Mar\'echal De Lattre De Tassigny, 75775 Paris cedex 16, France.\\
 Gosaef,  Facult\'{e} des Sciences de Tunis, 2092 Manar II-Tunis, Tunisia\\
 Email:  emmanuel.lepinette@ceremade.dauphine.fr}

%
%
%
%
\begin{abstract}
In a discrete time setting, we study the central problem of giving a fair price to some financial product. 
For several decades, the no-arbitrage conditions and the martingale measures have played a major role for solving this problem.  We propose a new approach for estimating the super-replication cost based on convex duality instead of martingale measures duality: The prices are expressed using Fenchel conjugate and bi-conjugate without using any no-arbitrage condition.The super-hedging problem resolution 
leads  endogenously to  a weak no-arbitrage condition  called Absence of Instantaneous Profit (AIP) under which prices are finite. We study this condition in details, propose several characterizations  and  compare it to the no-arbitrage condition. 
\end{abstract}


\end{frontmatter}

\section{Introduction}
Giving a fair price to a financial asset $G$ is a major question in the economic and financial theory. A selling price should be an amount which is enough to initiate a hedging strategy for $G$, i.e., a strategy whose value at maturity is always above $G$. It seems also natural to ask for the infimum of such amount. This is the so called  super-replication price and  it has been introduced in the binomial setup for transaction costs by \cite{BLPS}. 
Characterizing and computing the super-replication price has become one of the central issue in the mathematical finance theory, see, \cite{KSh}, \cite{FollS} and the references therein.  
It has mainly be addressed assuming that some no arbitrage condition holds true. Roughly speaking, this condition means that we cannot hope to make a profit without taking some risk. This condition has a good mathematical characterization in terms of existence of martingale measures, which is called the fundamental theorem of asset pricing (FTAP in short). 
The FTAP was initially formalised in \cite{HK79,HP81, K81},
while \cite{DMW} established it in a general discrete-time setting, and \cite{DS94} did so in continuous-time models. The literature on the subject is vast, and we refer to \cite{DelSch05, FollS} for a general overview. 
The FTAP is essential for pricing issues as it allows the characterisation of the super-replication price using the martingale measures. This is the so-called dual formulation of the superreplication price or superhedging theorem (see \cite{Schal99} and \cite{FK97} and the references therein).

In this paper, a discrete time model without frictions is considered where all the no-arbitrage conditions are equivalent. So, from now on, we call it   NA. Our goal is to compute a super-hedging (or super-replicating) prices of a European claim 
without assuming any normative condition such as NA on the financial market.  To do so, we  analyse from scratch the  set of super-hedging prices and its infimum value which will be called  the infimum super-hedging cost. Instead of the usual financial duality based on martingale measures under NA,  we compute the super-hedging cost using convex duality without postulating any  condition on the market.  Indeed, we express the one-step set of super-hedging prices  using Fenchel-Legendre conjugate and the infimum super-replication cost is obtained by the Fenchel-Legendre biconjugate.   To do so, we use  the notion of conditional essential supremum and, through measurable selection techniques, we prove  that the conditional essential supremum of a function of a random variable $Y$ is equal to the usual supremum of the function evaluated on  the conditional support of $Y$  (see Proposition \ref{lemma-essup-h(X)}).

The pricing formula that we obtain (see \eqref{eqbiconj})
shows that, if  $0$ does not belong to the convex hull of the conditional  support  of the price increment, 
then the super-hedging cost is equal to $-\infty$. To exclude this unrealistic possibility we postulate
the condition of Absence of Instantaneous Profit  (AIP). 
So AIP condition is indeed the minimal requirement in order to get a financial market where pricing is possible. 
The AIP is very weak~: If the initial information is trivial, a one period instantaneous profit is a strategy starting from $0$ and leading to a terminal wealth larger than some strictly positive constant. So the adjective instantaneous means that the gain is realized at time $0$. The AIP condition is easy to check in practice~: 
it suffices to verify that the cost of some non negative Call option is non negative. 

As the set of payoffs that may be super-replicated from $0$ is not closed under AIP, we  introduce an asymptotic version of the AIP condition called Absence of  Weak Instantaneous Profit  (AWIP) in the spirit of the No Free Lunch condition. We show that NA implies AWIP which implies AIP but that the reverse implications may not hold true. We also show that AWIP is  equivalent to the existence of absolutely continuous martingale measures.

Assuming AIP, we obtain that the one-step infimum super-hedging cost is finite and is the concave envelop of the payoff relatively to the convex envelop of the conditional support. Fenchel-Legendre duality have already been used to obtain a dual representation of the super-replication price thanks to deflators (see \cite[Exemple 4.2]{PenMOR} and \cite[Theorem 10 and Corollary 15]{PenMF}).  In \cite[Theorem 10]{PenMF}  the result is shown under the assumption that the set
of claims that can be super-replicated from $0$
is closed, which holds true under NA.   Here, our approach is different as we do not postulate any no-arbitrage assumptions on the market and  we do not seek for a dual representation of the super-hedging price. So, our main result shows that NA is not necessary to solve the super-hedging problem and that AIP is  the condition that ensures finite prices. Moreover,   we show that 
if a market satisfies AIP and is extended with a contingent claim priced with its super-replication cost, the extended market is still AIP free. 
This is not true changing AIP with NA as soon as the market is incomplete. 
Note that AIP condition is tailor made for pricing issues but is not sufficient to solve the problem of expected utility maximization, where  NA is required to obtain a well-posed problem.


%

The paper is organized as follows. In Section \ref{secone}, we study the one-period framework while in Section \ref{secmulti} we study the multi-period one.  
The proofs of technical results are postponed to the appendix.\smallskip

In the remaining part of this introduction, we present our framework and notations. Let $(\Omega,(\cF_t)_{t \in \{0,\ldots,T\}}\cF_T,P)$ be a complete filtered probability space,  where $T$ is the time horizon.
For any $\sigma$-algebra $\cH$ and any $k \geq 1$,  we denote by $L^0(\R^k,\cH)$ the set of $\cH$-measurable and $\R^k$-valued random variables.
We consider   a   non-negative process
$S:=\left\{S_{t},\ t \in \{0,\ldots,T\}\right\}$ such that  $S_{t} \in L^0(\R^d,\cF_t)$ for all $t \in \{0,\ldots,T\}$. The vector $S_t$ represents the  prices at time $t$ of  $d$ risky assets in the
financial market of consideration.
Trading
strategies are given by    processes $\theta:=\{ \theta_{t}, \, t \in \{0,\ldots,T-1\}\}$ such that  $\theta_{t} \in L^0(\R^d,\cF_{t})$ for all $t\in \{0,\ldots,T-1\}$. The vector $\theta_{t}$ represents the
investor's holding in   the $d$ risky assets between times $t$ and  $t+1$.
We assume that trading is self-financing and that the riskless asset's price is a constant equal to $1$. The value at time $t$ of a portfolio $\theta$ starting from
initial capital $x\in\mathbb{R}$ is then given by  \vskip -0.5cm

$$
V^{x,\theta}_t=x+\sum_{u=1}^t  \theta_{u-1} \Delta S_u,
$$
where $\Delta S_u=S_u-S_{u-1}$ for $u \geq 1$ and $xy$ is the scalar product  of $x$ and $y$.   
\section{The one-period framework}
\label{secone}
Let $\cH$ and $\cF$ be two complete sub-$\sigma$-algebras of $\cF_T$ such that $\cH \subseteq \cF$ and which represent respectively the initial and the final information. Let  $y\in L^0(\R^d,\cH)$   and $Y\in L^0(\R^d,\cF)$ be  two non-negative random variables.
They represent the initial and the final prices of the $d$ risky assets. We also consider a contingent claim $Z  \in L^0(\R,\cF)$. We will be particularly interested by derivatives on $Y$ i.e., $Z=g(Y)$ with $g: \Omega \times \R^d\to \R$ and  $g(Y):\o \mapsto g(Y)(\o)=g(\omega,Y(\omega)).$ \\
The objective of the section  is to obtain  a characterization of   $\cP(Z)$  the one-step set of super-hedging (or super-replicating) prices of $Z$ and of its infimum value. The setting will be applied in Section \ref{secmulti} with the choices $\cH=\cF_t$, $\cF=\cF_{t+1}$, $Y=S_{t+1}$ and $y=S_t$. 
\begin{defi}
\label{defpg}
The set $\cP(Z)$ of super-hedging prices of the contingent claim $Z\in L^0(\R,\cF)$ consists in  the initial values of super-hedging strategies $\theta$
$$\cP(Z)=\{x\in  L^0(\R,\cH),  \exists\, \theta\in L^0(\R^d,\cH),\;   x+\theta (Y -y)\ge Z \, {\rm a.s.}\}.$$
The infimum super-hedging cost of $Z$ is defined by $p(Z):=\essinf[]\cP(Z)$.\\
When $Z=g(Y)$ we write $\cP(g)=\cP(Z)$ and $p(g) = p(Z)$.  
\end{defi}

The notions of essential infimum and, more generally, conditional essential infimum, denoted by $\essinf[\cH],$ and conditional essential supremum,   denoted by $\esssup[\cH],$ are at the heart of this study and are defined in Proposition \ref{Essup}  below. We  also use   the notion of conditional support  of $Y,$ denoted by ${\rm supp}_{\cH}Y,$  which is introduced in Definition \ref{DefD} below.
In Section \ref{secgreve}, we derive the characterization of the super-hedging prices and cost from the following steps~:
\begin{enumerate}
\item Observe that the set of super-hedging prices can be rewritten using a conditional essential supremum (see  \eqref{eqpgZ} and \eqref{eqpg}).
\item Under mild conditions, show that the conditional essential supremum of a function of $Y$ is equal to the usual supremum of the function evaluated on the random set ${\rm supp}_{\cH}Y$ (see Proposition \ref{lemma-essup-h(X)}).
\item When $Z=g(Y)$, recognize that a super-hedging price can be written using a Fenchel-Legendre conjugate (see \eqref{eqf*}).
\item Take the essential infimum of the set of super-hedging prices and go through the steps 2. and 3. to recognize the Fenchel-Legendre biconjugate (see \eqref{eqbiconj}).
\item Use the classical convex biconjugate theorem   to evaluate the infimum super-hedging cost.
\end{enumerate}
With this pricing formula in hand (see \eqref{eqbiconj}) the Absence of Instantaneous Profit  (AIP) condition appears as the necessary and sufficient condition  to get  finite super-hedging costs. In Section \ref{secAIP}, we develop the concept of AIP and  propose several characterizations of the AIP condition. In Section  \ref{secComp} we compare it with the classical no-arbitrage  condition NA.

\subsection{Conditional support and conditional essential infimum}
\label{consupessinf}
This section is the toolbox of the paper. The proofs are postponed to the appendix.
We recall some results and notations that will be used without  further references in the rest of the paper.
 Let  $h: \Omega \times \R^d \to \R$.
The effective domain of $h(\o,\cdot)$ is defined by
$${\rm dom \,} h(\o,\cdot):=\{x \in \R^d, \, h(\o,x)< \infty\}$$ and $h(\o,\cdot)$ is proper if dom $h(\o,\cdot) \neq \emptyset$ and $h(\o,x)>- \infty$  for all $x \in \R^d$.
Next,   if $h$  is $ \cH$-normal integrand (see Definition 14.27 in \cite{rw}) then $h$ is $\cH \otimes \cB(\R^d)$-measurable and is lower semi-continuous (l.s.c. in the sequel, see \cite[Definition 1.5]{rw}) in $x$ and the converse holds true if $\cH$ is complete for some measure, see \cite[Corollary 14.34]{rw}.
Note that if $z \in L^0(\R^d,\cH)$ and $h$ is $\cH \otimes \cB(\R^d)$-measurable, then
$h(z) \in L^0(\R,\cH)$.\\
A random set $\cK : \Omega \twoheadrightarrow \mathbb{R}^{d}$ is  $\cH$-measurable if for all open set $O$ of $\R^d$, the subset $\{\o \in \O, \, O \cap \cK(\o) \neq \emptyset \} \in \cH$.
If $\cK$ is a $\cH$-measurable and closed-valued random set of $\R^d$, then $\cK$ admits a Castaing representation $(\eta_n)_{n\in \mathbb{N}}$ (see \cite[Theorem 14.5 ]{rw}). This means that
$\cK(\o)={\rm cl}\{\eta_n(\o),\,n\in \mathbf{N}\}$ for all $\o \in {\rm dom \, } \cK :=\{\o \in \Omega, \, \cK(\o)\cap \R^d \neq \emptyset\}$  where the closure is taken in $\R^d$.\smallskip

We introduce  the conditional support of $X \in L^0(\R^d,\cF)$ with respect to $\cH$.
\begin{defi}
\label{DefD}
Let $\mu$ be a $\cH$-stochastic kernel i.e., for all $\o \in \Omega$, $\mu(\cdot,\o)$ is a probability measure on $\cB(\R^d)$ and  $\mu(A,\cdot)$ is $\cH$-measurable for all $A \in \cB(\R^d)$. We define the random set  ${D}_{\mu} : \Omega \twoheadrightarrow \mathbb{R}^{d}$ by   \vskip -0.6cm
\begin{align}
\label{defd1}
{D}_{\mu}(\o):=\bigcap \left\{ A \subset \mathbb{R}^{d},\; \mbox{closed}, \; \mu(A,\o)=1\right\}.
\end{align}
For $\o \in {\O}$, ${D}_{\mu}(\o) \subset \mathbb{R}^d$ is called the support of $\mu(\cdot, \o)$. Let $X\in L^0(\R^d,\cF)$, we denote by $\supp[\cH]X$ the set defined in \eqref{defd1} when $\mu(A,\o)=P(X \in A | \mathcal{H})(\o)$ is a regular version of the conditional law of $X$ knowing $\cH$. The random set
 $\supp[\cH]X$ is called the conditional support of $X$ with respect to $\cH$.
\end{defi}
\begin{rem}
\label{remsupp}{\rm
When $\cH$ is the trivial sigma-algebra, $\supp[\cH]X$ is just  the usual support of $X$   (see p 441 of   \cite{ab}).
Theorems 12.7  and 12.14 of \cite{ab} show  that
$P(X \in . | \mathcal{H})$ admits a unique support $\supp[\cH]X\subset \mathbb{R}^d$ such that    $P(X \in \supp[\cH]X | \mathcal{H})=1$ a.s. i.e.,  ${\rm supp}_{\cH}X$ is a.s. non-empty.\\
For simplicity we will assume that $Y(\o) \in \supp[\cH]Y (\o)$ for all $\o \in \Omega$.
Moreover, as  $0\leq Y < \infty$, ${\rm  Dom \; supp \; }_{\cH}Y=\Omega$.}
\end{rem}
\begin{lemm}
\label{Dmeasurability} Let $\mu$ be as in Definition \ref{DefD}.
$D_{\mu}$ is non-empty, closed-valued and $\mathcal{H}$-measurable.
\end{lemm}

%
%

It is possible to incorporate measurability in the definition of the essential supremum (see \cite[Section 5.3.1]{KS}  for the definition and the proof of existence of the classical essential supremum). This has been done by  \cite{BCJ} for a single real-valued random variable and by \cite{KL} for a family of vector-valued random variables and with respect to a random partial order (see \cite[Definition 3.1 and Lemma 3.9]{KL}). Proposition \ref{Essup} is given and proved for sake of completeness and for pedagogical purpose. The authors thanks T. Jeulin who suggested this (elegant) proof.
\begin{prop}\label{Essup} Let $\cH$ and $\cF$ be complete $\sigma$-algebras such that $\cH\subseteq \cF$ and let $\Gamma=(\gamma_i)_{i\in I}$ be a family of real-valued $\cF$-measurable random variables. There exists a unique $\cH$-measurable random variable $\gamma_{\cH}\in L^0(\R \cup\{\infty\},\cH),$ denoted by $\esssup \Gamma,$ which satisfies the following properties
\begin{enumerate}
\item For every $i\in I$, $\gamma_{\cH}\ge \gamma_i$ a.s.
\item If $\zeta \in  L^0(\R \cup\{\infty\},\cH)$ satisfies  $\zeta\ge \gamma_i$ a.s. $\forall i\in I$, then $\zeta\ge \gamma_{\cH}$ a.s.
\end{enumerate}
\end{prop}

The conditional essential infimum $\essinf \Gamma$ is defined symmetrically. \smallskip 

{\sl Proof }{\it of Proposition \ref{Essup}.}
Considering the homeomorphism $\arctan$ we can restrict our-self to $\gamma_i$ taking values in $[0,1]$. We denote by $P_{\gamma_i|\cH}$ a regular version of the conditional law of $\gamma_i$ knowing $\cH$. Let $\zeta \in  L^0(\R \cup\{\infty\},\cH)$ such that $\zeta\ge \gamma_i$ a.s. $\forall i\in I$.
This is equivalent to $P_{\gamma_i|\cH}(]-\infty,x])|_{x=\zeta }=1 \mbox{ a.s. }$ and $\supp[\cH]\gamma_i \subset ]-\infty,\zeta ]$ a.s. follows from Definition \ref{DefD}.
Let
\begin{eqnarray}
\label{eqthierry}
\Lambda_{\gamma_i|\cH}=\sup\{x \in [0,1], \, x \in \supp[\cH]\gamma_i\}.
\end{eqnarray}
Then $\Lambda_{\gamma_i|\cH} \leq \zeta $ a.s.
and it  is easy to see that
$\Lambda_{\gamma_i|\cH}$ is $\cH$-measurable. So taking the classical essential supremum, we get that
$\mathrm{ess\,sup} \{ \Lambda_{\gamma_i|\cH},\, i \in I\} \leq \zeta $ a.s. and that $\mathrm{ess\,sup} \{ \Lambda_{\gamma_i|\cH},\, i \in I\}$ is $\cH$-measurable. We conclude that
$\gamma_{\cH}=\mathrm{ess\,sup} \{ \Lambda_{\gamma_i|\cH},\, i \in I\}$ a.s. since for every $i\in I$,
$P(\gamma_i\in  \supp[\cH]\gamma_i|\cH)=1$
(see Remark \ref{remsupp}). 
\fdem

\begin{lemm}
\label{lemsuppess} Assume that $d=1$ and consider $X\in L^0(\R,\cF)$. Then, we have a.s.  that
\begin{eqnarray}
\nonumber
\essinf X&=&\inf {\rm supp}_{\cH}X,\quad
\esssup X=\sup {\rm supp}_{\cH}X,\\
\nonumber
\essinf X&\in& {\rm supp}_{\cH}X \quad  {\rm on\,\, the\,\, set\,\,} \{\essinf X>-\infty\},\\
\nonumber
\esssup X&\in& {\rm supp}_{\cH}X \quad  {\rm on\,\, the\,\, set\,\,} \{\esssup X<\infty\},\\
\label{rein}
{\rm conv}{\rm supp}_{\cH}X & = & [\essinf X,\esssup X]\cap \R,
\end{eqnarray}
where ${\rm conv}{\rm supp}_{\cH}X$ is the convex envelop of ${\rm supp}_{\cH}X$ i.e., the smallest convex set that contains ${\rm supp}_{\cH}X$.
\end{lemm}

The following proposition is one of the main ingredient of the paper.
It extends the fact that $\esssup[\cH] X=\sup_{x\in \supp[\cH]X} x \mbox{ a.s.}$ (see \eqref{eqthierry}) and allows to compute a conditional essential supremum as a classical supremum but on a random set. A generalization is given in \cite{LM1}, see e.g. \cite{LM2}.
\begin{prop}\label{lemma-essup-h(X)} Let $X \in L^0(\R^d, \cF)$ be such that
${\rm dom \; } \supp[\cH]X =\Omega$
and let $h:$ $\Omega \times \R^d \to \R$
be a $\cH \otimes \cB(\R^d)$-measurable function which is  l.s.c.  in $x$.
Then,
\begin{eqnarray}
\label{belequa}
\esssup[\cH] h(X)=\sup_{x\in \supp[\cH]X} h(x) \quad a.s.
\end{eqnarray}
\end{prop}
%

\subsection{Fenchel-Legendre conjugate and bi-conjugate to express super-replication prices and cost}
\label{secgreve}
We are now in position to perform the program announced in the beginning of the section. 
Let  $Z \in  L^0(\R,\cF).$ As $x\in \cP(Z)$ if and only if there exists $\theta\in L^0(\R^d,\cH)$ such that   $x \ge Z-\theta (Y-y) \, {\rm a.s.}$, we get by definition of the conditional essential supremum (see Proposition \ref{Essup}) that 
\bea
\cP(Z) & =& \left\{\esssup[\cH]\left(Z-\theta (Y-y) \right),~\theta\in L^0(\R^d,\cH)\right\}+L^0(\R_+,\cH),\label{eqpgZ} \\
\label{eqpgZZ}
p(Z) & =& \essinf[]\left\{\esssup[\cH]\left(Z-\theta  (Y-y) \right),~\theta \in L^0(\R^d,\cH) \right\}.
\eea

In the case where $Z=g(Y)$ we are able to perform an explicit computation of $p(Z)$. To do so we recall that the (upper) closure $\overline{h}$ of $h$ is the smallest u.s.c.  function which dominates $h,$ i.e., 
$\overline{h}(x)=\limsup_{y \to x} h (y)$.  The lower closure is defined symmetrically.
\begin{theo}
\label{lempart1}
The set $\cP(g)$ of Definition \ref{defpg} can be express as follows
\begin{eqnarray}
\label{eqpg}
\cP(g)=\left\{\esssup[\cH]\left(g(Y)-\theta Y \right) + \theta y,~\theta\in L^0(\R^d,\cH)\right\}+L^0(\R_+,\cH).
\end{eqnarray}
Suppose that $g$ is a $\cH$-normal integrand.  Then, for $\theta \in L^0(\R^d,\cH)$, we get that
\begin{eqnarray}
\label{eqf*}
\esssup[\cH]\left(g(Y)-\theta Y \right)=\sup_{z\in {\rm supp}_{\cH}Y}\left(g(z)-\theta z \right)=f^*(-\theta ) \; \; \; {\rm a.s.}
 \end{eqnarray}
where $f$ and $f^*$, its Fenchel-Legendre conjugate, are given by
\begin{eqnarray}
\nonumber
\label{deff}
f(\o,z)& = &-g(\o,z)+\delta_{{\rm supp}_{\cH}Y(\o)}(\o,z)\\
f^*(\o,x) & = & \sup_{z\in \R^d}\left(xz -f(\o,z)\right),
\end{eqnarray}
and $\delta_{C(\o)}(\o,z)=0$ if $z\in C(\o)$ and $+\infty$ else. Moreover suppose that $g$ is proper and that there exists some concave function $\varphi$  such that $g\le \varphi<\infty$ on ${\rm conv}{\rm supp}_{\cH}Y$\footnote{This is equivalent to assume that there exists $\a, \b \in \R$, such that $g(x)\le \a x + \b $ for all $x \in {\rm conv}{\rm supp}_{\cH}Y$.}.
We have that a.s. 
 \begin{eqnarray}
 \label{eqbiconj}
p(g)& =& -f^{**}(y)  =   \overline{{\rm conc}}(g,{\rm supp}_{\cH}Y)(y)-\delta_{ {\rm conv}{\rm supp}_{\cH}Y}(y)\\
\nonumber
 & = & \inf{\{\a y + \b, \, \a \in \R^d,\,\b \in \R,\; \a x + \b \geq g(x),\, \forall x  \in  {\rm supp}_{\cH}Y\}}  -\delta_{ {\rm conv}{\rm supp}_{\cH}Y}(y),
\end{eqnarray}
where  $f^{**}$ is the Fenchel-Legendre biconjugate of $f,$ i.e.,
$f^{**}(\o,x)=\sup_{z\in \R^d}\left(xz -f^*(\o,z)\right)$ 
and the relative concave envelop of $g$ with respect to ${\rm supp}_{\cH}Y$ is given by
$$
{\rm conc}(g, {\rm supp}_{\cH}Y)(x)=
\inf\{ v(x),~ v\,{\rm\, is\, concave\, and\, }v(z)\ge g(z),\,\forall z\in {\rm supp}_{\cH}Y \}.$$
\end{theo}
Notice that the infimum super-hedging cost is not a priori a price, i.e., an element of $\cP(g)$, as the later may be an open interval. Note also that  \cite{CGT} and \cite{BN14} have represented the super-hedging price as
a concave envelop  but this was formulated under the no-arbitrage condition using the dual representation of the super-replication price  through martingale measures.  
\begin{rem}
\label{teemu}{\rm
Fenchel-Legendre duality have already been used many times in financial mathematics. In particular, Pennanen obtains a dual representation of the super-replication price thanks to deflators (see \cite[Example 4.2]{PenMOR} and \cite[Theorem 10 and Corollary 15]{PenMF}).  The proof of \cite[Theorem 10]{PenMF} is also based on the (convex) biconjugate theorem but the result is shown under the assumption that the set $\cR$
of claims that can be super-replicated from $0$ (see \eqref{argenttropcher})
is closed, which holds true under the no-arbitrage condition NA.  In \cite{PenPerkMP}, the existence and the absence of duality gap in a general stochastic optimization problem is proved through dynamic programming and under a condition (that does not rely on inf-compactness) of linearity on sets constructed with recession functions. This condition in classical mathematical finance problems is equivalent to the no-arbitrage condition (see \cite[Exemple 1]{PenPerkMP}). Our approach is different as we do not postulate any assumption on the market and we obtain from the biconjugate representation a formula for the infimum super-hedging cost (see \eqref{eqbiconj}). We then deduce the condition under which the market prices are finite (see Proposition \ref{NGDone}). Importantly, our goal is not to obtain a dual representation thanks to deflator or martingale measures.}
\end{rem}
{\sl Proof of Theorem \ref{lempart1}.}
First \eqref{eqpg}  follows from \eqref{eqpgZ}.  
Lemma \ref{Dmeasurability} will be in force. Under
the assumption that $g$ is a $\cH$-normal integrand,  \eqref{eqf*}  follows from Proposition \ref{lemma-essup-h(X)}. Under the additionnal assumptions that $g$ is proper and that there exists some concave function $\varphi$  such that $g\le \varphi<\infty$ on ${\rm conv}{\rm supp}_{\cH}Y$, we show first that ${\rm conv \,}f$ satisfies \eqref{convconc} below and is  proper, where ${\rm conv\,} f$ is the convex envelop of $f$ i.e., the greatest convex function dominated by $f.$
It can be written as follows  (see \cite[Proposition  2.31]{rw})
\begin{eqnarray*} {\rm conv \,}f(x)   =&  \inf\left\{\sum_{i=1}^n \l_i f(x_i), \,  n \geq 1,\,  (\l_i)_{i\in \{1,\ldots,n\}} \in \R_+^n, \,  (x_i)_{i\in \{1,\ldots,n\}} \in \R^{d \times n}, \right. \\
 & \left. x=\sum_{i=1}^n \l_i  x_i, \, \sum_{i=1}^n \l_i=1\right\}.
 \end{eqnarray*}
Let $x=\sum_{i=1}^n \l_i x_i$ for some $ n \geq 1,\,  (\l_i)_{i\in \{1,\ldots,n\}} \in \R_+^n$ such that $\sum_{i=1}^n \l_i=1$ and  $(x_i)_{_{i\in \{1,\ldots,n\}}} \in \R^{d \times n}$.
Assume that $x \notin {\rm conv}{\rm supp}_{\cH}Y$. Then (see \cite[Proposition  2.27, Theorem 2.29]{rw}),
there exists at least one $x_i \notin {\rm supp}_{\cH}Y$ and $f(x_i)=+\infty$ and also ${\rm conv \,}f(x)=+\infty$.
If  $x \in {\rm conv}{\rm supp}_{\cH}Y$,  ${\rm conv \,}f(x)=-{\rm conc}(g, {\rm supp}_{\cH}Y)(x)$ by definition.
Moreover, for  all $x\in {\rm conv}{\rm supp}_{\cH}Y$, $x=\sum_{i=1}^n \l_i x_i$, with  $ n \geq 1$ and $ (\l_i)_{i\in \{1,\ldots,n\}} \in \R_+^n$, and  $(x_i)_{i\in \{1,\ldots,n\}} \subset {\rm supp}_{\cH}Y$ such that $\sum_{i=1}^n \l_i=1$, we have
$${{\rm conc}}(g, {\rm supp}_{\cH}Y)(x)\geq \sum_{i=1}^n \l_i {{\rm conc}}(g, {\rm supp}_{\cH}Y)(x_i) \geq \sum_{i=1}^n \l_i g(x_i)>-\infty.$$ Moreover, ${\rm conc}(g, {\rm supp}_{\cH}Y)\le \varphi <\infty \mbox{ on }{\rm conv}{\rm supp}_{\cH}Y.$
Thus,   for all $ x\in {\rm conv}{\rm supp}_{\cH}Y$, ${\rm conc}(g, {\rm supp}_{\cH}Y)(x) \in \R$
and one may write that \begin{eqnarray}
\label{convconc}
{\rm conv \,}f &=&-{\rm conc}(g, {\rm supp}_{\cH}Y) +\delta_{{\rm conv}{\rm supp}_{\cH}Y} \; \; \; {\rm a.s.}\end{eqnarray}
As   ${\rm conv}{\rm supp}_{\cH}Y$ is non-empty,  ${\rm conv \,}f$ is  proper and
\cite[Theorem 11.1]{rw} implies that  $f^*$ is proper, l.s.c and convex and that
$
f^{**}(y)= \underline{{\rm conv}} \,f(y).$
Moreover, using Lemma \ref{mesmes}, $f^*(\o,x)=\sup_{z \in {\rm supp}_{\cH}Y(\o)}\left( xz + g(\o,z)\right)$ is $\cH\otimes \mathcal{B}(\R^d)$-measurable. We obtain that a.s.
 \begin{eqnarray*}
p(g)& =&\essinf[]\{f^*(-\theta)+ \theta y,~\theta\in L^0(\R^d,\cH)\}=
 -\esssup[]\{\theta y - f^*(\theta),~\theta\in L^0(\R^d,\cH)\},
\end{eqnarray*}
using \eqref{eqpg}. 
Assume for a moment that 
\begin{eqnarray}
\label{lala}
\esssup[]\{\theta y - f^*(\theta),~\theta\in L^0(\R^d,\cH)\}= \sup_{z\in \mathbb{R}^d }\left(z y-f^*(z)\right) \mbox{ a.s.}
\end{eqnarray}
is proved. Then \eqref{convconc} implies that 
\begin{eqnarray*}
p(g)    &=&-f^{**}(y)=-\underline{{\rm conv}} \,f(y)
={\overline{\rm conc}}(g, {\rm supp}_{\cH}Y)(y) - \delta_{{\rm conv}{\rm supp}_{\cH}Y}(y) \; \; \; {\rm a.s.}
\end{eqnarray*}
and the equality below \eqref{eqbiconj} follows easily. 

For any $z\in \R^d$, $\esssup[]\{\theta y - f^*(\theta),~\theta\in L^0(\R^d,\cH)\}\geq z y - f^*(z)$ a.s.
and $
\esssup[]\{\theta y - f^*(\theta) ,~\theta\in L^0(\R^d,\cH)\}\geq  \sup_{z\in \mathbb{R}^d }\left(z y-f^*(z)\right) \mbox{ a.s.}$ follows. \\
Conversely, for all $\theta\in L^0(\R^d,\cH)$, we have $\theta y - f^*(\theta) \leq \sup_{z\in \R^d }\left(z y-f^*(z)\right)$ a.s. 
 If $\sup_{z\in \R^d }\left(z y-f^*(z)\right)$ is $\cH$-measurable,  \eqref{lala} holds true. 
As $f^*$ is  $\cH \otimes \cB(\R^d)$-measurable, 
${\rm graph\,}{\rm dom \;}f^*=\{(\omega,x)\in \Omega\times \R^d, \; f^*(\omega,x)<\infty\} \in \cH \otimes \cB(\R^d)$ and
${\rm dom \;}f^* $ is $\cH$-measurable (see  \cite[Theorem  14.8]{rw}).
As $f^*$ is proper, ${\rm dom} f^*(\o,\cdot) \neq \emptyset$.
By measurable selection argument, we obtain the existence of a $\cH$-measurable selector $a$ such that $a\in {\rm dom}(f^*)$ a.s. 

On ${\rm dom}(f^*)=\{a\}$, we have $\sup_{z\in \R^d }\left(z y-f^*(z)\right)=a y-f^*(a)$ which   is $\cH$-measurable.
Otherwise,
on ${\rm dom}(f^*)\ne \{a\}$, ${\rm ri \,}{\rm dom}(f^*) \ne \emptyset$ and  $z \mapsto z y-f^*(z)$ is concave hence continuous on this set and   we show below that a.s. 
\begin{eqnarray}
\label{eqlast}
\sup_{z\in \R^d }\left(z y-f^*(z)\right) & = &
\sup_{z\in{\rm ri \,}{\rm dom} f^* }\left(z y-f^*(z)\right)= \sup_{z\in \hat\Gamma  }\left(z y-f^*(z)\right)
 \end{eqnarray}
 where $\hat\Gamma $ is a countable $\cH$-measurable dense subset of $\R^d$. So $\sup_{z\in \R^d }\left(z y-f^*(z)\right)$ is $\cH$-measurable, which allows to conclude. 
Let us prove \eqref{eqlast}. The first equality is classical (see for example  Lemma A.32 of \cite{BC}). 
 As ${\rm dom}(f^*)$ is $\cH$-measurable so is its affine hull $H$ (see \cite[Exercice  14.12]{rw}). Let $(q_n)_{n\ge 1}$ be a dense subset of $\R^d$ and let us denote by $p_{H}(q_n)$ the projection of $q_n$ onto $H$, $n\ge 1$. By  \cite[Exercice  14.17]{rw},  $p_{H}(q_n)$ is $\cH$-measurable for all $n\ge 1$ and it is well known that $(p_{H}(q_n))_{n\ge 1}$ is dense in $H$. 
So  any $z\in {\rm ri \,}{\rm dom}(f^*) \subseteq H$ is the limit of a countable (with random index) family $(z_n)_{n\ge 1}$, which is a subfamily of the larger countable family $\hat\Gamma(\cdot)=(p_{H}(q_n)(\cdot))_{n\ge 1}$. By continuity on the relative interior, we then deduce that $\sup_{z\in{\rm ri \,}{\rm dom} f^* }\left(z y-f^*(z)\right)\le \sup_{z\in \hat\Gamma  }\left(z y-f^*(z)\right)$ hence the second equality in \eqref{eqlast} holds by the first one.  \fdem

\subsection{The AIP condition}
\label{secAIP}
Theorem \ref{lempart1} shows that if $ y\notin {\rm conv}{\rm supp}_{\cH}Y$ the infimum super-hedging cost of a European claim $p(g)$ equals $-\infty$. Avoiding this situation leads to the notion of absence of instantaneous profit that we present now.  Let $\cR$ be the set of all $\cF$-measurable claims that can be super-replicate from $0$:
\begin{eqnarray}
\label{argenttropcher}
\cR:=\left\{ \theta(Y-y)-\epsilon^+,~\theta \in L^0(\R^d,\cH),\;\epsilon^+\in L^0(\R_+,\cF)\right\}.
\end{eqnarray}
Then
\begin{eqnarray*}
\cP(0) =  \{x\in  L^0(\R,\cH),  \exists\, \theta\in L^0(\R^d,\cH),\;   x+\theta (Y -y)\ge 0 \, {\rm a.s.}\}
=  (-\cR) \cap L^0(\R,\cH).
\end{eqnarray*}
Note that $0\in  \cP(0),$ so $p(0) \leq 0.$
We say that there is an instantaneous profit when $P(p(0)<0)>0$ i.e., if it is possible to super-replicate the contingent claim $0$ at a negative super-hedging price.
 \begin{defi}
 \label{defipone}
 There is an  instantaneous profit (IP)  if
 $P(p(0)<0)>0$.  On the contrary case  if  $p(0)=0$ a.s. we say that the Absence of Instantaneous Profit (AIP) condition holds.
 \end{defi}
 \vskip -0.2cm


We know propose several characterisations of the AIP condition. 
\begin{prop}
\label{NGDone}
AIP holds  if and only if one of the following condition holds true.
\begin{enumerate}
\item  $y \in {\rm conv}{\rm supp}_{\cH}Y$ a.s.  or $0 \in {\rm conv}{\rm supp}_{\cH}(Y-y)$ a.s.
\item $\sigma_{{\rm supp}_{\cH}(Y-y)} \geq 0$ a.s. where $\sigma_{D}(z)=\sup_{x \in D} (-xz)$ is the support function of $-D$
\item $\cP(0) \cap L^0(\R_-,\cH)=\{0\}$  or $\cR \cap L^0(\R_+,\cH)=\{0\}.$
\end{enumerate}
\end{prop}
By the third statement above, we see that $X$ is an instantaneous profit if $X\in  L^0(\R_+,\cH) \setminus \{0\}$ and if there exists some 
$\theta \in L^0(\R^d,\cH)$ such that $\theta (Y-y)\ge X$. The term instantaneous means that the profit is realized at $t=0$ since $X$ is $\cH$-measurable, see also Remark \ref{remoa2}. 
\begin{rem}
\label{foi}
{\rm
In the case $d=1$,  \eqref{rein} implies that the previous conditions are equivalent to
$y \in  [\essinf[\cH] Y, \esssup[\cH]  Y] \cap \R\, {\rm a.s.}$
}
\end{rem}
\begin{rem}
\label{remoa2}
{\rm
The AIP condition is tailor-made for pricing purposes. It allows to give a finite super-hedging cost even in case of arbitrage opportunity (see Example \ref{calc} below).
Note that an IP is a very strong strategy. Assume that $\cH$ is trivial, then an IP corresponds to some $\theta \in \R^d$ such that $\theta(Y-y)\ge c$ for some deterministic  constant $c>0$. Set $p_0=-c$, $p_0$ is a super-hedging price that allows to get the zero payoff at time $1$, i.e., $p_0+\theta (Y-y)\ge 0$ a.s. 
It is clear that $2p_0$ is still a super-hedging price for the zero claim using the strategy $2\theta$. As 
$$p_0+2\theta (Y-y)=2p_0+2\theta (Y-y)+(-p_0)\ge c,$$ 
it is indeed possible to get a terminal portfolio value larger than a deterministic strictly positive quantity.
}
\vskip -1cm

\end{rem}


{\sl Proof of Proposition \ref{NGDone}.}
The assumptions of Theorem \ref{lempart1} are satisfied for $g=0$
and we get that
$p(0)=-\delta_{{\rm conv}{\rm supp}_{\cH}Y}(y)$ a.s. Hence, AIP holds true if and only if $ y\in  {\rm conv}{\rm supp}_{\cH}Y$ a.s. or equivalently $0 \in {\rm conv}{\rm supp}_{\cH}(Y-y)$ a.s.  and AIP is equivalent to 1.
Using Theorem \ref {lempart1}, we get that
\begin{eqnarray*}
\cP(0) & =& \left\{\esssup[\cH]\left(-\theta (Y-y) \right),~\theta\in L^0(\R^d,\cH)\right\}+L^0(\R_+,\cH).
\end{eqnarray*}
Proposition \ref{lemma-essup-h(X)} implies that for $\theta\in L^0(\R^d,\cH)$,
$$\esssup[\cH]\left(-\theta (Y-y) \right)=\sup_{x \in  {\rm supp}_{\cH}(Y-y)}\left(-\theta x \right)=\sigma_{{\rm supp}_{\cH}(Y-y)}(\theta).$$
So, $\cP(0) \cap L^0(\R_-,\cH)=\{0\}$  if and only if $\sigma_{{\rm supp}_{\cH}(Y-y)} \geq 0$ a.s. and $2.$ and $3.$ are equivalent. To finish the proof, it remains to prove that  2. is equivalent to
1.  First remark that 
$\sigma_{{\rm supp}_{\cH}(Y-y)}= \sigma_{{\rm conv}{\rm supp}_{\cH}(Y-y)}.$
So, it remains to prove that, for any  closed convex set $D$ of $\R^d$,  $\sigma_D \geq 0$ if and only if $0 \in D$.  If  $0 \in D$ it is clear that
$\sigma_D \geq 0$. Assume that $0 \notin D$. Then, by Hahn-Banach theorem, there exists some $\b>0$ and some $\theta_0 \in \R^d \setminus\{0 \}$ such that
$-x \theta_0 \leq -\beta$ for all $x \in D$ and $\sigma_D(\theta_0) \leq -\beta<0$ follows.
\fdem
\begin{coro}
\label{aipcall} The AIP condition holds true if and only if $p(g) \geq 0$ a.s.  for some  non-negative $\cH$-normal integrand $g$ such that there exists some concave function $\varphi$  verifying that $g\le \varphi < \infty$.
\end{coro}
In particular, the AIP condition holds true if and only if the infimum super-hedging cost of some European call option is non-negative. Note that, under AIP, the price of some non-zero payoff call option may be zero, see Example \ref{calc} below and \cite{Ruf}.\smallskip

{\sl Proof of Corollary \ref{aipcall}.}
Assume AIP  holds true. Then, by Definition \ref{defipone}, we get that $p(0)=0$ a.s. As $g \geq 0$, it is clear that $p(g) \geq p(0)=0$ a.s.
Conversely, assume that there exists an IP and let $g$ be some non-negative $\cH$-normal integrand  such that there exists some concave function $\varphi$  verifying that $g\le \varphi < \infty$,  
Proposition \ref{NGDone} leads to  $P(y \in {\rm conv}{\rm supp}_{\cH}Y)<1$ and, since 
$\overline{{\rm conc}}(g,{\rm supp}_{\cH}Y)(y) \leq \varphi<\infty,$ \eqref{eqbiconj} implies that $P(p(g)=-\infty)>0.$ The converse is proved. \fdem \vskip -1cm

\subsection{Comparison between AIP and NA}
\label{secComp}
\vskip -0.2cm

 \begin{defi}
 \label{defiNA}
The no-arbitrage (NA) condition holds true if $\theta(Y- y) \geq 0$ a.s. for some $\theta \in L^0(\R^d,\cH)$ implies that $\theta(Y-y) = 0$ a.s. or equivalently   $ \cR\cap L^0(\R_+,\cF)=\{0\}.$
 \end{defi}
 
Recall that in discrete time financial models without frictions, NA is equivalent to 
other definitions of no-arbitrage used in the literature: the No Unbounded Profit with Bounded Risk (NUPBR),  the No Arbitrage  of First Kind  (NA1), 
the No free Lunch with Vanishing Risk (NFLVR) and the No Unbounded Increasing Profit (NUIP), see  \cite{DelSch05},  \cite{KKS}, \cite{KK},  \cite{Font} and  Remark 2.3 in \cite{FR1}. So, we only have to compare AIP to NA. 

 \begin{lemm} 
\label{NUPBR} The   NA condition  implies AIP and the reverse may not hold true. 
\end{lemm}
{\sl Proof }{\it of Lemma \ref{NUPBR}.}  Assume that there exists an instantaneous profit i.e., some $X\in  L^0(\R_+,\cH)$ such that $P(X>0)>0$ and some 
$\theta \in L^0(\R^d,\cH)$ such that $\theta (Y-y)\ge X$. Then, the strategy $\theta$ leads clearly to an arbitrage opportunity according to NA.  On the contrary, AIP may hold while NA fails. 
Fix $d=1$, $\cH=\{\emptyset,\Omega\}$, $y=0$ and $Y$ follows an uniform distribution on $[0,1]$. Then
$P(Y-y>0)=1$ and the  constant strategy equal to 1 leads to an arbitrage opportunity. Nevertheless $y=0 \in {\rm conv}{\rm supp}_{\cH}Y=[0,1]$ and AIP holds true. \fdem \smallskip

We now provide a necessary and sufficient condition for the equivalence between AIP and NA.
\begin{prop}
\label{NAAIP}
AIP  and NA are equivalent if and only if
\begin{eqnarray}
\label{jacshiCNS}
0  \in {\rm conv}{\rm supp}_{\cH}(Y-y) \mbox{ a.s.} \Leftrightarrow
0\in {\rm ri } \,({\rm conv}{\rm supp}_{\cH}(Y-y))\mbox{ a.s.}
\end{eqnarray}
This last condition is satisfied if
\begin{eqnarray}
\label{jacshi}
P(0  \notin {\rm conv}{\rm supp}_{\cH}(Y-y) \setminus {\rm ri } \,({\rm conv}{\rm supp}_{\cH}(Y-y)))=1.
\end{eqnarray}
\end{prop}
\noindent {\sl Proof of Proposition \ref{NAAIP}.} 
Proposition \ref{NGDone} shows that AIP is equivalent to $0 \in {\rm conv}{\rm supp}_{\cH}(Y-y)$ a.s. On the other hand, \citep[Theorem 3g)]{JS98}  shows that NA  is equivalent to $0 \in {\rm ri } \,({\rm supp}_{\cH}(Y-y))$ a.s.
So AIP is equivalent to NA if and only if \eqref{jacshiCNS} holds true. This last condition is implied by \eqref{jacshi}.
\fdem 
\begin{rem}{\rm
The proof  of Proposition \ref{NAAIP} when $d=1$ enlightens the difference between AIP and NA and is given for pedagogical purpose. Assume that AIP holds true and that  $P(\essinf[\cH]  Y=y)=P(\esssup[\cH]Y=y)=0$. 
Using Remark  \ref{foi},
$y \in  [\essinf[\cH] Y, \esssup[\cH]  Y] \cap \R\, {\rm a.s.}$
Let $\theta \in L^0(\R,\cH)$ such that  $\theta(Y-y)\ge 0$. On the set $\{\theta>0\}\in \cH$, we have that $Y\ge y$ hence $\essinf[\cH] Y\ge y \geq \essinf[\cH] Y$. We deduce that $P(\theta>0)=0$. Similarly, we get that $P(\theta<0)=0$. Finally $\theta=0$, NA holds true and the proposition is proved as 
$${\rm conv}{\rm supp}_{\cH}(Y-y) \setminus {\rm ri } \,({\rm conv}{\rm supp}_{\cH}(Y-y))=\{\essinf[\cH]  Y-y, \esssup[\cH]  Y-y\}.$$ 
}
\end{rem}

We finish this section with the following interesting result: Contrary to NA,  AIP makes it possible to obtain super-hedging cost which are AIP free. 
We adopt the definition and concepts of \cite[Section 1]{FollS}. 
Let $Z\in L^0(\R,\cF)$ be a  fixed contingent claim and let $p(Z)$ be defined in Definition \ref{defpg}. 
We show that if the initial market $(y,Y)$, satisfying AIP, is extended  with $(p(Z),Z),$ we obtain a market which is still free of instantaneous profits. We say that $p(Z)$ is  instantaneous profit-free, in the spirit of \cite[Definition 1.30]{FollS} where the concept was introduced for the no-arbitrage condition NA and the super-replication price. This problem naturally arises in the models under NA but also, more generally, for other types of no-arbitrage condition, as in \cite{SH}.

As in Definition \ref{defipone},  AIP in the extended market holds true if $p^{Y,Z}(0)=0$ where $p^{Y,Z}(0)$ is the super-hedging cost of $0$ in the extended market, i.e., 
\bean &&p^{Y,Z}(0)=\\
&& \essinf[]\left\{x\in  L^0(\R,\cH),  \exists\, \alpha \in  L^0(\R,\cH),\, \theta\in L^0(\R^d,\cH),\;  x+\alpha (Z-p(Z))+\theta (Y-y )\ge 0 \, {\rm a.s.} \right\}.\eean

\begin{theo} 
\label{AIPext}
The AIP condition holds if and only if it holds true in the extended market, i.e., the market with the additional  asset $(p(Z),Z)$.
\end{theo}
{\sl Proof of Theorem \ref{AIPext}.} 
If the extended market satisfies AIP, then Proposition \ref{NGDone} implies that $0_{\R^{2}} \in   {\rm conv}{\rm supp}_{\cH}  (\hat Y - \hat y)\, {\rm a.s.}$  where $\hat y=(y,p(Z))$ and 
$\hat Y=(Y,Z)$. As 
\begin{eqnarray}
\label{eqsuppproduit}
{\rm supp}_{\cH}(\hat Y - \hat y) \subset {\rm supp}_{\cH}( Y-y) \times {\rm supp}_{\cH}( Z -p(Z)),
\end{eqnarray}  
$0 \in   {\rm conv}{\rm supp}_{\cH}(Y-y)\, {\rm a.s.}$ and 
the initial market satisfies AIP as well. Reciprocally, suppose that the initial market satisfies AIP. Consider a super-hedging price $x\in L^0(\R,\cH)$ in the extended market for the zero claim at time $1,$ i.e., such that $x+\alpha (Z-p(Z))+\theta (Y-y )\ge 0$ a.s. where $\alpha \in  L^0(\R,\cH)$ and $\theta \in L^0(\R^d,\cH)$. We show below that $x\ge 0$ a.s. which will imply that $p^{Y,Z}(0)=0$ and thus AIP holds in the extended market. Let $A^1=\{\alpha<0\}\in \cH$. Then, a.s. 
$$\left(\frac{x}{-\alpha}+p(Z)\right)1_{A^1}+\frac{\theta 1_{A^1}}{-\alpha}(Y-y)\ge Z1_{A^1}.$$
Let $\bar x \in \cP(Z )$ and $\bar \theta \in  L^0(\R^d,\cH)$ such that $\bar x+ \bar \theta (Y-y) \geq Z$ a.s. 
Then,  a.s. 
$$\left( \bar x  1_{\Omega \setminus A^1} + \left(\frac{x}{-\alpha}+p(Z) \right)1_{A^1}\right)+
\left(\bar \theta  1_{\Omega \setminus A^1}+  \frac{\theta 1_{A^1}}{-\alpha} \right)(Y-y)\ge Z,$$
$ \bar x1_{\Omega \setminus A^1} +\left(\frac{x}{-\alpha}+p(Z) \right) 1_{A^1} \geq p(Z)$ and 
 we conclude that $x\ge 0$ on $\{\alpha<0\}$.

On the set $A^2=\{\alpha=0\}\in \cH$, we have $x 1_{A^2}+\theta 1_{A^2}(Y-y)\ge 0$ a.s.  Therefore, 
$x 1_{A^2} \geq p(0)=0$ 
since AIP holds true for the initial market defined only by $(y,Y)$.  We deduce that $ x \ge 0$ on $\{\alpha=0\}$.

At last, before analysing the problem on the set $A^3=\{\alpha>0\}\in \cH$, let us recall that, as $\cP(Z)$ is 
downward-directed, there exists $z_n \in \cP(Z)$, for all $n$ such that 
$p(Z)=\lim_n \downarrow z^n$. 
Fix $n>0$. Let $k_n=\inf\{k, \, z^k <p(Z)+n^{-1}\}$. Then,  by a measurable selection argument one may assume that $k_n \in L^0(\R,\cH).$ Let 
$r^n=\sum_{l \geq 0} z^l 1_{\{k_n=l\}}$. As $\{k_n=l\} \in \cH$ and $z^l \in \cP(Z),$ $r^n\in \cP(Z)$ and   $r^n+\theta^n(Y-y)\ge Z$  a.s. for some $\theta^n\in L^0(\R^d,\cH)$.  Hence  $p(Z)\le r^n\le p(Z)+n^{-1}$ and  $Z-p(Z)\le \theta^n(Y-y)+n^{-1}$. Therefore, on the set $A^3$, we have that  a.s. 
\bean  \frac{x}{\alpha}+n^{-1}+\left(\theta^n+\frac{\theta}{\alpha}\right)(Y-y)\ge \frac{x}{\alpha}+(Z-p(Z))+\frac{\theta}{\alpha}(Y-y)\ge 0\eean
and $(\frac{x}{\alpha}+n^{-1})1_{A^3} \geq p(0)=0$ as AIP holds for the initial market defined by $(y,Y)$. 
Therefore, when $n\to \infty$, we deduce that $x\ge 0$ on $\{\alpha>0\}$. The conclusion follows. \fdem

\begin{coro} 
\label{corocoherent}
Suppose that AIP holds. Then, $p(Z) \in [ \essinf[\cH]  Z,\esssup[\cH]  Z] \cap \R\, {\rm a.s.}$
\end{coro}
{\sl Proof } {\it of Corollary \ref{corocoherent}.} Suppose that AIP holds true. Then the extended market i.e., the market  with the additional  asset $(p(Z),Z)$ satisfies the AIP condition by the theorem above. Thus \eqref{eqsuppproduit} shows that 
$0 \in   {\rm conv}{\rm supp}_{\cH}(  Z -p(Z))\, {\rm a.s.}$ 
and Remark \ref{foi} implies that $p(Z) \in [ \essinf[\cH]  Z,\esssup[\cH]  Z] \cap \R\, {\rm a.s.}$ \fdem

\begin{rem}{\rm Observe that Theorem \ref{AIPext} does not hold true in an incomplete market for the super-replication price and NA.  Indeed, let us consider a one step incomplete market. Let $Z\in L^0(\R,\cF)$ be a non replicable contingent claim. Then \cite[Theorem 1.32]{FollS} implies that there exists $\theta \in \R^d$ such that
$z+\theta (Y-y)\ge Z$ a.s. where  
$z=\inf\{x \in \R, \, \exists \theta \in \R^d, \, x + \theta  (Y-y)\ge Z \mbox{ a.s.}\}$ is the  super-replication  price of $Z$. As $Z$ is not replicable $P(z+\theta (Y-y)> Z)>0$ and 
$\theta(Y-y) - (Z-z)$ is a.s. non negative and strictly positive with strictly positive probability. Therefore, NA fails in the extended market  $((y,Y),(z,Z)).$ }
\end{rem}

\subsection{Super-hedging cost under AIP} 
We now provide the characterization of the infimum super-hedging cost under the AIP condition.
\begin{coro}
\label{fenchngd} Suppose that AIP holds true. Let $g$ be a proper $\cH$-normal integrand such that there exists some concave function $\varphi$  verifying that $g\le \varphi<\infty$ on  ${\rm conv}{\rm supp}_{\cH}Y$. Then, a.s.
\begin{equation}\label{eqfenchprix}
\begin{split}
p(g)&=\overline{{\rm conc}}(g, {\rm supp}_{\cH}Y)(y)\\
 &=\inf{\{\a y + \b, \, \a \in \R^d,\,\b \in \R,\;  \a x + \b\geq g(x),\, \forall x\in {\rm supp}_{\cH}Y\}}.\end{split}
\end{equation}
If $g$ is concave and u.s.c., $p(g)=g(y)$ a.s.
\end{coro}
{\sl Proof } {\it of Corollary \ref{fenchngd}.} 
The two equalities are direct consequence of Theorem \ref{lempart1}. If $g$ is concave and u.s.c., the result is trivial. \fdem  \smallskip

We finish the one-period analysis with the computation of the infimum super-hedging cost of a convex derivative when $d=1$.
In this case, the cost is in fact a super-hedging price and we get the super-hedging strategy explicitly.
\begin{coro}
\label{fenchngdconv} Suppose that AIP holds true and that $d=1$. Let $g:\R \to \R$ be a non-negative convex function with ${\rm dom} \, g=\R$ and $\lim_{x\to \infty} x^{-1}g(x)=M \in [0,\infty)$, then a.s.
\begin{eqnarray}\label{PricingFormula}
p(g) & = &  \theta^* y + \beta^*=g(\essinf[\cH] Y ) + \theta^*\left( y-\essinf[\cH]Y \right),\\
 \label{theta}
\theta^* & = & \frac{g(\esssup[\cH] Y )-g(\essinf[\cH]Y)}{\esssup[\cH] Y-\essinf[\cH] Y},
\end{eqnarray}
where we use the conventions $\theta^*=\frac{0}{0}=0$ in the case $\esssup[\cH] Y=\essinf[\cH] Y$ a.s. and $\theta^*=\frac{g(\infty)}{\infty}=M$ if $\essinf[\cH] Y<\esssup[\cH] Y=+\infty$ a.s. Moreover, $p(g)\in \cP(g)$.
\end{coro}
{\sl Proof} {\it of Corollary \ref{fenchngdconv}.} 
As $g$ is convex, the relative concave envelop of $g$ with respect to ${\rm supp}_{\cH}Y$ is the affine function that coincides with $g$ on the extreme points of the interval ${\rm conv}{\rm supp}_{\cH}Y$ and \eqref{PricingFormula} and \eqref{theta} follow from \eqref{eqfenchprix} and Remark \ref{foi}. Then using \eqref{eqfenchprix} and \eqref{PricingFormula}, we get that (recall that $Y \in {\rm supp}_{\cH}Y$)
\begin{eqnarray}\label{lavieestbelle}
p(g)+ \theta^* (Y-y) =\theta^* Y + \beta^* \geq g(Y) \; {\rm a.s.}
\end{eqnarray}
and $p(g)\in \cP(g)$ follows.
\fdem 
\begin{ex}
\label{calc}{\rm
We compute the price of a call option under AIP in the case $d=1$. Let $G=g(Y)=(Y-K)_+$ for some $K\geq 0$.
\begin{itemize}
\item If $K\geq \esssup[\cH] Y$ then $Y-K\leq \esssup[\cH] Y -K$ and $G=0$. As AIP condition holds true, $p(g)=p(0)=0$.
\item If $K\leq \essinf[\cH] Y$ then $Y-K\geq \essinf[\cH] Y -K$ and $G=Y-K$. As $g$ is concave and u.s.c., $p(g)=g(y)= y-K$ a.s.
\item If $\essinf[\cH] Y\leq K \leq \esssup[\cH] Y.$ Then, \eqref{theta} and \eqref{PricingFormula} imply that
\begin{eqnarray*}
p(g) & = &  \frac{\esssup[\cH] Y -K }{\esssup[\cH] Y-\essinf[\cH] Y}\left( y-\essinf[\cH]Y \right)
\end{eqnarray*}
on $\{\esssup[\cH] Y\neq \essinf[\cH] Y\}$ and 0 else.
So $p(g)=0$ if and only if  $y=\essinf[\cH]Y$ or $\esssup[\cH] Y= \essinf[\cH] Y$. A non-negative call option can have a zero price, see also \cite{Ruf}.
\end{itemize}
We finish with an example of computation for a call price under AIP when NA fails.  
We assume  a.s.  that for $y>0$,  $\essinf[\cH]Y=dy$ and  $\esssup[\cH]Y=u y$ for two constants $u$ and $d$. By Remark \ref{foi},
AIP is equivalent to $d\leq 1 \leq u$. If ($d=1$ and $u>1$) or ($u=1$ and $ d<1$),   AIP holds true but NA fails.   Suppose that $d=1$ and $u>1$. If $K \geq y$, the super-replication price under AIP is zero and if $K \leq y$, it is $y-K$. The same holds true if $u=1$ and $ d<1$. }
\end{ex}

\section{The multi-period framework}
\label{secmulti}
\subsection{Multi-period super-hedging prices}
For every $t \in \{0,\ldots,T\},$ the set $\mathcal{R}_t^T$ of all  claims that can be super-replicated from the zero initial endowment at time $t$ is defined by
\begin{eqnarray}
\label{defR}
\mathcal{R}_t^T:=\left\{\sum_{u=t+1}^T \theta_{u-1}\Delta S_u-\epsilon_T^+,~\theta_{u-1}\in L^0(\R^d,\cF_{u-1}),\;\epsilon_T^+\in L^0(\R_+,\cF_{T})\right\}. \quad
\end{eqnarray}
The set of (multi-period) super-hedging prices and the (multi-period) infimum super-hedging cost  of some contingent claim $g_T\in L^0(\R,\cF_T)$ at time $t$  are given for all $t \in \{0,\ldots,T\},$ by
\begin{eqnarray}
\nonumber
\cP_{T,T}(g_T)& = & \{g_T\} \mbox{ and }
\p_{T,T}(g_T)  = g_T\\
\label{defPi}
\cP_{t,T}(g_T)   &= &  {\{x_t\in L^0(\R,\cF_{t}),\, \exists R \in \mathcal{R}_t^T,\, x_t+R=g_T  \mbox{ a.s.}\}} \\
\nonumber
\p_{t,T}(g_T) & = &\essinf[]\cP_{t,T}(g_T).
\end{eqnarray}
As in the one-period case, it is clear that the infimum super-hedging cost is not necessarily a price in the sense that $\p_{t,T}(g_T)\notin  \cP_{t,T}(g_T)$ when $ \cP_{t,T}(g_T)$ is not closed. 

We now define a local version of super-hedging prices. The set of one-step super-hedging prices of the payoff $g_{t+1} \in  L^0(\R,\cF_{t+1})$  and it associated infimum super-hedging cost are given by
\bea \nonumber
\cP_{t,t+1}(g_{t+1}) & =& \left\{x_t\in  L^0(\R,\cF_{t}),  \exists\, \theta_t\in L^0(\R^d,\cF_{t}),\;   x_t+\theta_t  \Delta S_{t+1} \ge g_{t+1} \, {\rm a.s.} \right\}\\\nonumber
\pi_{t,t+1}(g_{t+1}) & =& \essinf[] \cP_{t,t+1}(g_{t+1})\\ \label{GeneralClaimPricing}
&=&\essinf[]\left\{\esssup[\cF_t]\left(g_{t+1}-\theta_t  \Delta S_{t+1} \right), \,\theta_t\in L^0(\R^d,\cF_{t})  \right\},
\eea
see \eqref{eqpgZZ}. In the following, we extend the definition of  $\cP_{t,u}(g_{u})$, $u\ge t +1$, so that the argument $g_u$ may be a subset $G_u\subseteq L^0(\R,\cF_u)$. Precisely, we set $\cP_{t,u}(G_{u})=\cup_{g_u\in G_u} \cP_{t,u}(g_{u})$.  The following lemma makes the link between local and global super-hedging prices. It also provides a dynamic programming principle, meaning that the prices are time consistent.
\begin{lemm}
\label{lemouf}
Let $g_T \in L^0(\R,\cF_{T})$ and $t \in \{0,\ldots,T-1\}$. Then
$$\cP_{t,T}(g_T)
=\cP_{t,t+1}(\cP_{t+1,T}(g_T)) \mbox{ and }
\pi_{t,T}(g_T) \geq \pi_{t,t+1}(\pi_{t+1,T}(g_T)).$$
Moreover, assume that
$\pi_{t+1,T}(g_T) \in \cP_{t+1,T}(g_T)$. Then $$\cP_{t,T}(g_T)=\cP_{t,t+1}(\pi_{t+1,T}(g_T)) \mbox{ and }
\pi_{t,T}(g_T)=\pi_{t,t+1}(\pi_{t+1,T}(g_T)).$$
\end{lemm}
{\sl Proof} {\it of Lemma \ref{lemouf}.} Consider $x_t\in \cP_{t,T}(g_T)$. Then, for all $u\in \{t,\ldots,T-1\}$,  there exist $\theta_{u}\in L^0(\R^d,\cF_{u})$   such that 
$$x_t+\theta_{t} \Delta S_{t+1}+\sum_{u=t+2}^{T}  \theta_{u-1} \Delta S_u  \geq g_T \mbox{ a.s.}$$ 
We deduce that $x_{t+1}:=x_t+\theta_{t} \Delta S_{t+1}\in \cP_{t+1,T}(g_T).$ Moreover $x_t\in \cP_{t,t+1}(x_{t+1})$ and $x_t\in \cP_{t,t+1}(\cP_{t+1,T}(g_T))$. Reciprocally, suppose that $x_t\in \cP_{t,t+1}(\cP_{t+1,T}(g_T))$, i.e., $x_t\in \cP_{t,t+1}(x_{t+1})$ for some $x_{t+1}\in \cP_{t+1,T}(g_T)$. Then, $x_t+\theta_{t} \Delta S_{t+1}\ge x_{t+1}$ for some $\theta_{t}\in L^0(\R^d,\cF_{t})$ and $x_{t+1}+\sum_{u=t+2}^{T}  \theta_{u-1} \Delta S_u  \geq g_T$ where $\theta_{u}\in L^0(\R^d,\cF_{u})$  for all $u\ge t+1$. It follows that 
$x_t+\sum_{u=t+1}^{T}  \theta_{u-1} \Delta S_u  \geq g_T \mbox{ a.s.}$ 
and $x_t \in \cP_{t,T}(g_T)$. 
Let $x_t \in \cP_{t,T}(g_T)
=\cP_{t,t+1}(\cP_{t+1,T}(g_T))$, then there exists $\theta_t \in L^0(\R^d,\cF_{t})$ and $x_{t+1}\in  \cP_{t+1,T}(g_T)$ such that 
$$x_t+ \theta_t  \Delta S_{t+1} \geq x_{t+1} \geq \essinf[]\cP_{t+1,T}(g_T)=\pi_{t+1,T}(g_T) \mbox{ a.s.}$$
Thus $x_t \in \cP_{t,t+1}(\pi_{t+1,T}(g_T) )$ and $\cP_{t,T}(g_T) \subset \cP_{t,t+1}(\pi_{t+1,T}(g_T) )$. 
Moreover 
 $x_t \geq \pi_{t,t+1}(\pi_{t+1,T}(g_T))$ and the first statement follows. 
If $\pi_{t+1,T}(g_T) \in \cP_{t+1,T}(g_T)$, then 
$\cP_{t,t+1}(\pi_{t+1,T}(g_T))  \subset \cP_{t,t+1}(\cP_{t+1,T}(g_T))=\cP_{t,T}(g_T)$. 
 \fdem

 \begin{rem}
{\rm 
Under AIP, if at each step, $\pi_{t+1,T}(g_T) \in \cP_{t+1,T}(g_T)$ and if we have that $\pi_{t+1,T}(g_T)=g_{t+1}(S_{t+1})$ for some ``nice'' $\cF_{t}$-normal integrand $g_{t+1}$, we will get from Corollary \ref{fenchngd} that $\pi_{t,T}(g_T)=\overline{{\rm conc}}(g_{t+1}, {\rm supp}_{\cF_{t}}S_{t+1})(S_{t})$ a.s., see Remark \ref{seexpli} for a tangible example. 

Note that the super-hedging problem is solved for general claims $\xi_T$  through the formula (\ref{GeneralClaimPricing}). For claims of  Asian type $g((S_u)_{u\le T})$ or of American type,  what we propose for European claims could be easily adapted. Consider a general claim $\xi_T$ and the natural  filtration,  i.e., the one generated by the price process $S$.  Then for any self-financing portfolio $\theta$,  $V_T= x+ \sum_{t=1}^T \theta_{t-1}\Delta S_{t} \ge \xi_T$ if and only if $V_T\ge\tilde \xi_T$ where $\tilde \xi_T={\rm esssup}_{\cF_T}\xi_T$. Thus $\xi_T$ and  $\tilde \xi_T$ have the same super-replication cost and as $\tilde \xi_T$ is $\cF_T$-measurable,  it is of the form $\tilde \xi_T=g((S_u)_{u\le T})$. Of course, in practice, it is necessary to have an idea about $g$ but the same difficulty arises under the NA condition. 
}
\end{rem}
\subsection{Multi-period AIP}

\begin{defi}
\label{foire}
The AIP  condition holds true if  for all  $t \in \{0,\ldots,T\}$
$$\cP_{t,T}(0)\cap L^0(\R_-,\cF_t)=\{0\}.$$
\end{defi}

We now study the link between global and local instantaneous profits. A global (resp. local) instantaneous profit means that it is possible to super-replicate from a negative cost at  time $t$ the claim $0$ paid at time $T$ (resp. time $t+1$).  The next proposition shows that the local and global AIP conditions are equivalent in the following sense.

\begin{prop}\label{thoNip} The following assertions are equivalent. 
\begin{enumerate}
\item $\cP_{t,T}(0)\cap L^0(\R_-,\cF_t)=\{0\}$ for all $t\in \{0,\ldots,T-1\},$ i.e., AIP.
\item $\cP_{t,t+1}(0)\cap L^0(\R_-,\cF_t)=\{0\}$ for all $t\in \{0,\ldots,T-1\}$.
\item $S_t \in {\rm conv}{\rm supp}_{\cF_t}S_{t+1}$  (or
$0 \in {\rm conv}{\rm supp}_{\cF_t}\Delta S_{t+1}$) a.s.  for all $t\in \{0,\ldots,T-1\}$.
\item $\s_{{\rm supp}_{\cF_t} \Delta S_{t+1}} \geq 0\, {\rm a.s.}$  for all $t\in \{0,\ldots,T-1\}$.
\item $\p_{t,T}(0)=0$ \, {\rm a.s.}  for all $t\in \{0,\ldots,T-1\}$.
\end{enumerate}
\end{prop}

{\sl Proof }{\it of Proposition \ref{thoNip}.}  
For some fixed $t\in \{0,\ldots,T-1\}$ we show that 
\begin{eqnarray}
\label{greve}
\cP_{t,T}(0)\cap L^0(\R_-,\cF_t)=\{0\} \Longleftrightarrow \p_{t,T}(0)=0.
 \end{eqnarray} 
For the implication, note that $\p_{t,T}(0) \leq 0$ is always true. Let $x_t \in  \cP_{t,T}(0)$. 
Then there exist $\theta_{u}\in L^0(\R^d,\cF_{u})$ for $u \geq t$ such that 
$x_t + \sum_{u=t+1}^T \theta_{u-1}\Delta S_u \geq 0$ a.s. Thus 
$x_t 1_{ x_t <0} + \sum_{u=t+1}^T \theta_{u-1}1_{ x_t <0}\Delta S_u \geq 0$ a.s. 
If $P(x_t <0)>0$ then $x_t 1_{ x_t <0} \in \cP_{t,T}(0)\cap L^0(\R_-,\cF_t)$, a contradiction. Thus $x_t \geq 0$ a.s. and $\p_{t,T}(0) =\essinf[]\cP_{t,T}(0) \geq 0.$ For the reverse implication  let $x_t \in \cP_{t,T}(0)\cap L^0(\R_-,\cF_t)$.  If $P(x_t <0)>0$ then $x_t 1_{ x_t <0} \in \cP_{t,T}(0)$ and $x_t 1_{ x_t <0} \geq \p_{t,T}(0)=0$ a.s., a contradiction. Thus  $x_t \geq 0$ a.s. and  $x_t = 0$ a.s. follows. \\
It is clear that \eqref{greve} implies that 1. is equivalent to 5. Now  we show that 1. is equivalent to 2. 
Suppose that 1. holds true. Then,  $\p_{t+1,T}(0)=0 \in \cP_{t+1,T}(0)$ and Lemma \ref{lemouf} implies that $\cP_{t,T}(0)  =  
\cP_{t,t+1}(\p_{t+1,T}(0))=\cP_{t,t+1}(0).$ 
and 2. holds. Reciprocally suppose that 2. holds true. Then, $\cP_{T-1,T}(0)\cap L^0(\R_-,\cF_{T-1})=\{0\}$. 
From \eqref{greve} with $t=T-1$ we get that $\p_{T-1,T}(0)=0 \in \cP_{T-1,T}(0)$
and  Lemma \ref{lemouf} implies that
$$\cP_{T-2,T}(0)=\cP_{T-2,T-1}(\p_{T-1,T}(0))=\cP_{T-2,T-1}(0)=L^0([0,\infty),\cF_{T-2}).$$
It is trivial that $ L^0([0,\infty),\cF_{T-2}) \subset  \cP_{T-2,T-1}(0)$.  The same reasoning as in the proof of the implication in \eqref{greve} with  $\cP_{T-2,T-1}(0)\cap L^0(\R_-,\cF_{T-2})=\{0\}$ proves the reverse inclusion. 
It follows that $\cP_{T-2,T}(0)\cap L^0(\R_-,\cF_{T-2})=\{0\}$. Using backward induction, 1. holds true. By Proposition \ref{NGDone} and Definition \ref{defipone}, we conclude that  3. and 4. are equivalent to 2.
\fdem

\subsection{Comparison with the NA condition}
\label{secomp}
 We first recall the classical multiperiod no-arbitrage (NA) condition.
 \begin{defi}
 \label{defiNAmulti}
 The no-arbitrage  (NA) condition holds  true if   for all $t \in \{0,\ldots, T\},$
$$\mathcal{R}_t^T \cap L^0(\R_+,\cF_T)=\{0\}.$$
 \end{defi}
It is easy to see that NA  can also be formulated as follows~:   $V_T^{0,\theta} \geq 0$ a.s. implies that $V_T^{0,\theta}= 0$ a.s. Recall that the set of all super-hedging  prices for the zero claim at time $t$ is given by $\cP_{t,T}(0)=(-\mathcal{R}_t^T)\cap L^0(\R,\cF_t)$ (see \eqref{defR} and \eqref{defPi}). It follows that (see Definition \ref{foire})
\bea \mbox{AIP holds true} \Leftrightarrow \mathcal{R}_t^T \cap L^0(\R_+,\cF_t)=\{0\} \mbox{ for all } t \in \{0,\ldots, T\}.\label{NA}\eea
It is clear that NA implies AIP and, as already mentioned, the equivalence does not hold true:
The AIP condition is strictly weaker than the  NA one.  

\begin{rem}\label{seexpli}
{\rm
It is possible to obtain the same computation schemes as in Proposition 2.2 of \cite{CV} assuming only AIP and not NA. Suppose   $\essinf[\cF_{t-1}]S_t=k^d_{t-1}S_{t-1}$ {\rm a.s.} and $\esssup[\cF_{t-1}]S_t=k^u_{t-1}S_{t-1}$ {\rm a.s.} where $S_0$, $(k^d_{t-1})_{t\in \{1,\ldots,T\}}$ and $(k^u_{t-1})_{t\in \{1,\ldots,T\}}$ are deterministic non-negative numbers. Then, AIP  holds true if and only if  $k^d_{t-1}\in [0, 1]$ and $k^u_{t-1}\in [1, +\infty]$ for all $t\in \{1,\ldots,T\}.$ Assume that AIP holds true, let $h$ be a convex function and let $H=h(S_T)$ be some European contingent claim. Then, the infimum super-hedging cost of  $H$ is  the replication price of $H$ in the binomial model where  $S_{t}\in \{k^d_{t-1}S_{t-1},k^u_{t-1}S_{t-1}\}$ a.s., for all $t\in \{1,\ldots,T\}$. 
This is proved in the companion paper \cite{BCL} where we also give  some promising numerical illustrations. Indeed, we  calibrate  historical data of the  French index CAC $40$ to this model and  implement the super-hedging strategy for a call option. Our procedure, which is  model free and based only on statistical estimations, provides better results that the one based on implied volatility. 
}
\end{rem}

As in the one period case, we are able to prove that the super-hedging cost is  instantaneous profit-free i.e., does not create instantaneous profits in the extended dynamic market where it is possible to trade the additional asset. 
In the following, $C_T\in L^0(\R,\cF_T)$ is fixed and  
$C_t=\pi_{t,t+1}(C_{t+1})$ is defined recursively for  $t\le T-1$ by \eqref{GeneralClaimPricing}. 
\begin{theo} 
\label{AIPextmulti}
The AIP condition holds true  if and only if the extended market, i.e., the market with the additional  asset $(C_t)_{t \in \{0,\cdots,T\}}$ satisfies AIP. \\
Suppose that AIP holds true. Then, $\essinf[\cF_{t}]  C_{t+1}\le C_t \le \esssup[\cF_{t}]  C_{t+1}\, {\rm a.s.}$ for all $t\in \{0,\ldots,T-1\}$. 
\end{theo}
{\sl Proof }{\it of Theorem \ref{AIPextmulti}.}
Proposition \ref{thoNip} implies that AIP holds true if and only if $\cP_{t,t+1}(0)\cap L^0(\R_-,\cF_t)=\{0\},$ for all $t\in \{0,\ldots,T-1\}$ and 
Theorem \ref{AIPext} shows that this last condition is equivalent to  $\pi_{t,t+1}^{S,C}(0)=0$ where $\pi_{t,t+1}^{S,C}(0)$ is the super-hedging cost in the extended market of $0$, i.e., the  essential infimum of all the 
$p_t\in L^0(\R,\cF_t)$  such that $p_t+\alpha_t\Delta C_{t+1}+\theta_t\Delta S_{t+1}\ge 0$ a.s. where $\alpha_t\in  L^0(\R,\cF_t)$ and $\theta_t\in L^0(\R^d,\cF_t)$. Using again Proposition \ref{thoNip} (in the extended market) this is equivalent to AIP in the extended market.  The last assertion follows from Corollary \ref{corocoherent}.
\fdem \smallskip

We now introduce  an asymptotic version of the AIP condition in the spirit of the No Free Lunch condition (NFL), i.e.,
considering the closure of the set $\mathcal{R}_t^T$ in (\ref{NA}).

 \begin{defi}
 \label{defiAWIPi}
The absence of weak instantaneous profit  (AWIP) condition holds true if for all $\in \{0,\ldots,T\}$
$$\overline{\mathcal{R}_t^T} \cap L^0(\R_+,\cF_t)=\{0\},$$  where the closure of $\mathcal{R}_t^T$ is taken with respect to the convergence in probability.
 \end{defi}

Recall that under NA, $\mathcal{R}_t^T$ is closed and that NFL and NA are equivalent. Under AIP, $\mathcal{R}_t^T$ may not be  closed and 
 we show in Lemma \ref{lemstrict} below that NA implies AWIP, which in turn implies AIP, but that the reverse implications may not hold true. 

Before, in the case $d=1$, we provide conditions under which   AWIP is equivalent to AIP. We also provide a characterization of AWIP through (absolutely continuous) martingale measures.
\begin{theo}\label{theo-EqWNFL} Assume that $d=1$. The following statements are equivalent.
\begin{enumerate}
\item AWIP  holds true.
\item For every $t \in \{0,\ldots,T\}$, there exists $Q\ll P$ with $\E(dQ/dP|\cF_t)=1$ such that $(S_u)_{u\in \{t,\ldots, T\}}$ is a $Q$-martingale.
\item AIP holds and $\overline{\mathcal{R}_t^T} \cap L^0(\R,\cF_t)=\mathcal{R}_t^T \cap L^0(\R,\cF_t)$ for every $t \in \{0,\ldots,T\}.$
\end{enumerate}
If $P(\essinf[\cF_t]S_{t+1}=S_t)=P(\esssup[\cF_t]S_{t+1}=S_t)=0$ for all $t\in \{0\ldots,T-1\}$ then AWIP, AIP and NA are equivalent. 
\end{theo}
The proof, which is postponed to the appendix, is based on classical Hahn-Banach Theorem arguments, see for example the textbooks of \cite{DelSch05} and  \cite{KS}.
\begin{rem}\label{propEqWNFL-AIP}{\rm
Theorem \ref{theo-EqWNFL} shows that  AIP and AWIP are equivalent if  $\mathcal{R}_t^T$ is  closed. Therefore, Lemma \ref{lemstrict} implies that $\mathcal{R}_t^T$ is  not necessarily closed under AIP, which is a key point in the classical theory under NA to obtain dual characterization of super-hedging prices. 
}
\end{rem}

\begin{lemm} \label{lemstrict} The AIP condition is not necessarily equivalent to AWIP and AWIP is not necessarily equivalent to NA.
\end{lemm}
{\sl Proof} {\it of Lemma \ref{lemstrict}.}
Assume that $d=1$.  Let us consider a positive process $(\tilde S_t)_{t\in \{0,\ldots,T\}}$ which is a $P$-martingale. We suppose that $\essinf[\cF_0]\tilde S_1<\tilde S_1$ a.s., which holds in particular
if  $\tilde S$ a geometric Brownian motion as $\essinf[\cF_0]\tilde S_1=0$ a.s.  Let us define $S_t:=\tilde S_t$ for $t\in\{1,\ldots,T\}$ and $S_0:=\essinf[\cF_0]S_1$. We have $\essinf[\cF_0]S_1\le S_0$ and $\esssup[\cF_0]S_1\ge \essinf[\cF_0]S_1=S_0.$ Hence AIP holds true at time $0$ (see Remark \ref{foi}). Moreover, by the martingale property (see Theorem \ref{theo-EqWNFL}), AIP and also AWIP hold at any time $t\in\{1,\ldots,T\}$. Let us suppose that AWIP  holds true at $t=0$. Using Theorem \ref{theo-EqWNFL}, there exists $\rho_T\ge 0$ with $\E(\rho_T)=1$ such that $S$ is a $Q$-martingale where $dQ=\rho_TdP$. Therefore, $\E(\rho_T\Delta S_1)=0$. Since $\Delta S_1>0$ by assumption, we deduce that $\rho_T=0$ hence a contradiction. 

Let us consider a one step model where $T=1$, $S_0=1$ and  $S_1$ is a random variable such that $S_1\ge 1$ a.s. and $P(S_1=1)\in (0,1)$. Let us define $Q^1$ by $dQ^1/dP=1_{\{S_1=1\}}/P(S_1=1)$. Then, $Q^1\ll P$ and $E_{Q^1}(S_1)=S_0.$ So  $(S_t)_{t\in \{0,1\}}$ is a $Q^1$-martingale and  AWIP holds true by Theorem \ref{theo-EqWNFL}. 
As $S_1-S_0 \geq 0$ a.s. and $P(S_1-S_0>0)>0$, NA fails. \fdem \smallskip

\section{Appendix}
\label{secappendix}
{\sl Proof of Lemma \ref{Dmeasurability}}.
It is clear from \eqref{defd1} that for all $\o \in \O$,  $D_{\mu}(\o)$ is a non-empty and closed subset of $\mathbb{R}^{d}$. We show that $D_{\mu}$ is  $\mathcal{H}$-measurable. Let $O$ be a fixed open set in $\mathbb{R}^{d}$ and
$
\mu_{O}: \omega \in \O \mapsto \mu_{O}(\o) :=  \mu(O,\o).
$
As $\mu$ is a stochastic kernel,  $\mu_{O}$ is $\mathcal{H}$-measurable.
By definition of  $D_{\mu}(\o)$ we get that $\{\o \in \Omega,\; {D}_{\mu}(\o) \cap O \neq \emptyset \} =\{\o \in \Omega, \; \mu_{O}(\omega)>0\} \in \mathcal{H}$ 
and $D_{\mu}$ is $\mathcal{H}$-measurable. \fdem \bigskip

\noindent {\sl Proof of Lemma \ref{lemsuppess}.} The two first statements follow from the construction of $\esssup X$ in Proposition \ref{Essup} (see \eqref{eqthierry}). Suppose that $\essinf X\notin {\rm supp}_{\cH}X$ on some non-null measure subset $\Lambda\in \cH$ of $\{\essinf X>-\infty\}$. As $\supp[\cH]X$ is $\cH$-measurable and closed-valued,  by a measurable selection argument, we deduce the existence of $r\in L^0(\R_+,\cH)$ such that $r>0$ a.s. and $(\essinf X-r, \essinf X+r)\subseteq \R\setminus {\rm supp}_{\cH}X$ on $\Lambda$. As $X\in {\rm supp}_{\cH}X$ a.s.  (see Remark \ref{remsupp}) and $X \geq \essinf X$ a.s., we deduce that $X\ge \essinf X+r$ on $\Lambda$, which contradicts the definition of  $\essinf X$. The next statement is similarly shown and the last one follows. \fdem \\

\noindent The proof of Proposition \ref{lemma-essup-h(X)} is based on the  two following useful lemmata.
\begin{lemm}\label{mes} Let $\cK:$ $\Omega \twoheadrightarrow \mathbb{R}^{d}$ be a $\cH$-measurable and closed-valued random set such that ${\rm dom \; }\cK=\O$
and  let $h:$ $\Omega \times \R^d \to \R$ be  l.s.c.  in $x$.  Then,
\begin{eqnarray}
\label{belequa2}\sup_{x\in \cK} h(x)=\sup_{n \in \mathbf{N}} h(\eta_n),
\end{eqnarray}
where $(\eta_n)_{n \in \mathbf{N}}$ is a Castaing representation of $\cK$.
\end{lemm}
{\sl Proof of Lemma \ref{mes}.} Let $\o  \in \O$.
As $(\eta_n(\o))_{n \in \mathbf{N}}\subset \cK(\o)$, $h(\o,\eta_n(\o)) \leq \sup_{x\in \cK(\o)} h(\o,x)$ and thus $\sup_n h(\eta_n) \leq \sup_{x\in \cK} h(x)$.
Let $x \in\cK(\o)={\rm cl}\{\eta_n(\o), \,n\in \mathbf{N}\}$, by lower semicontinuity of $h$, we get that 
$h(\o,x) \leq \liminf_n h(\o,\eta_n(\o))\le \sup_n h(\o,\eta_n(\o)).$ 
We conclude that  $\sup_{x\in \cK} h(x) \leq \sup_n h(\eta_n)$ and \eqref{belequa2} is proved.
\fdem

\begin{lemm}\label{mesmes} Let $\cK:$ $\Omega \twoheadrightarrow \mathbb{R}^{d}$ be a $\cH$-measurable and closed-valued random set such that ${\rm dom \; }\cK=\O$
and let $h:$ $\Omega \times \R^k \times \R^d \to \R$ be a $\cH \otimes \cB(\R^k)\otimes \cB(\R^d)$-measurable function such that
  $h(\o,x,\cdot)$  is  l.s.c.  for all
$(\o,x) \in \Omega \times \R^k $. Then
 $(\o,x) \in \Omega \times \R^k  \mapsto s(\o,x)=\sup_{z\in \cK(\o)} h(\o,x,z)$ is $\cH \otimes \cB(\R^k)$-measurable. \end{lemm}
{\sl Proof of Lemma \ref{mesmes}.}
Lemma \ref{mes}  implies that
$
s(\o,x)=\sup_n h(\o,x,\eta_n(\o)),
$
where $(\eta_n)_{n \in \mathbf{N}}$ is a Castaing representation of $\cK$.
So  for any fixed  $c \in \R,$ we get that 
\begin{eqnarray*}
\{(\omega,x) \in \O \times \mathbb{R}^{d}, \, s(\o,x)\le c \}
&=&\cap_n\{(\omega,x)\in \O \times \mathbb{R}^{d}, \, h(\o,x,\eta_n(\omega))\le c \}.
\end{eqnarray*}
As $h$ is $\cH \otimes \cB(\R^k)\otimes \cB(\R^d)$-measurable and  $\eta_n$ is $\cH$-measurable, $(\omega,x) \mapsto h(\o,x,\eta_n(\omega))$ is $\cH \otimes \cB(\R^k)$-measurable and so is $s$.
\fdem\\

\noindent {\sl Proof of Proposition \ref{lemma-essup-h(X)}.}
As $P(X \in  \supp[\cH]X|\cH)=1$ (see Remark \ref{remsupp}) we have that
$\sup_{x\in \supp[\cH]X} h(x) \geq h(X)$ a.s. and the  definition of
$\esssup[\cH] h(X)$ implies that  $\sup_{x\in \supp[\cH]X} h(x) \geq \esssup[\cH] h(X)$ a.s. since $\sup_{x\in \supp[\cH]X} h(x)$ is $\cH$-measurable by  Lemmata \ref{Dmeasurability} and \ref{mesmes}.

Let $(\gamma_n)_{n \in \mathbf{N}}$ be a   Castaing representation of $\supp[\cH]X.$ Lemmata \ref{Dmeasurability} and \ref{mes} imply that
$\sup_{x\in \supp[\cH]X} h(x)=\sup_n h(\gamma_n).$
Fix some rational number $\e>0$ and some integer $n>0$ and set $Z_{\e,n}= 1_{B(\gamma_n,\e)}(X)$, where $B(\gamma_n,\e)$ is the closed ball of center $\gamma_n$ and radius $\e$.
Let $\Omega_{\e,n}=\{E(Z_{\e,n}|\cH)>0\}.$ Then $P(\Omega_{\e,n})=1$. Otherwise, $P(\Omega \setminus \Omega_{\e,n})>0$ and on $\Omega \setminus \Omega_{\e,n} \in \cH$,
$P(X\in \R^d \setminus B(\gamma_n,\e)|\cH)=1$ and by definition \ref{DefD}, $\supp[\cH]X \subset \R^d \setminus B(\gamma_n,\e)$, which contradicts $\gamma_n\in \supp[\cH]X.$
By definition of the conditional essential supremum, we have that $\esssup[\cH] h(X)\geq h(X)$ a.s. and that $\esssup[\cH] h(X)$ is $\cH$-measurable. This implies that, for all fixed $\o \in \O_{\e,n}$,
\begin{eqnarray*}
\esssup[\cH] h(X)(\o)& \geq &  \frac{\E(Z_{\e,n}h(X)|\cH)}{\E(Z_{\e,n}|\cH)}(\o) =   \frac{\int 1_{B(\gamma_n(\o),\e)}(x)h(\o,x)  P_{X|\cH}(dx;\o)}{\E(Z_{\e,n}|\cH)(\o)} \\
& \geq   &   \frac{\int \left(\inf_{y \in B(\gamma_n(\o),\e)}h(\o,y)\right)1_{B(\gamma_n(\o),\e)}(x) P_{X|\cH}(dx;\o)}{\E(Z_{\e,n}|\cH)(\o)}\\
 & \geq & \inf_{y \in B(\gamma_n(\o),\e)}h(\o,y).
\end{eqnarray*}
As  $h$ is l.s.c. (recall \cite[Definition 1.5, equation 1(2)]{rw}), we have that
$$\lim_{\e \to 0} \inf_{y \in B(\gamma_n,\e)} h(y)=\liminf_{x \to \gamma_n}h(x)=h(\gamma_n).$$
So on the full measure set $\cap_{\e \in \mathbb{Q}\,\e>0, n \in \mathbb{N}} \O_{e,n}$,
$\esssup[\cH] h(X)\geq h(\gamma_n)$. Taking the supremum over all $n$, we get that
$$\esssup[\cH] h(X) \geq \sup_n h(\gamma_n)=\sup_{x\in \supp[\cH]X} h(x) \geq \esssup[\cH] h(X) \mbox{ a.s. \fdem}$$

\noindent {\sl Proof of Theorem \ref{theo-EqWNFL}.} First we prove that 1. implies 2. Suppose that AWIP  holds and fix some $t\in \{0,\ldots, T\}$. We may suppose without loss of generality that the process  $S$ is integrable under $P$. Under AWIP, we then have $\overline{\mathcal{R}_t^T} \cap L^1(\R_+,\cF_t)=\{0\}$ where the closure is taken in $L^1$. Therefore,   for every nonzero $x\in L^1(\R_+,\cF_t)$, there exists by the Hahn-Banach theorem a non-zero $Z_x\in L^{\infty}(\R_+,\cF_T)$ such that (recall that $\mathcal{R}_t^T$ is a cone) $\E Z_xx>0$ and  $\E Z_x\xi\le 0$ for every $\xi\in  \mathcal{R}_t^T$. Since $-L^1(\R_+,\cF_T)\subseteq \mathcal{R}_t^T$, we deduce that $Z_x\ge 0$ and we way renormalise $Z_x$ so that $\| Z_x\|_{\infty}=1$.  Let us consider the family
$$\cG=\{\{ \E(Z_x|\cF_t)>0\},~x\in L^1(\R_+,\cF_t)\setminus \{0\}\}.$$ Consider any non-null set $\Gamma\in \cF_t$. Taking $x=1_{\Gamma}\in L^1(\R_+,\cF_t)\setminus \{0\}$, since $\E(Z_x1_{\Gamma})>0$, we deduce that $\Gamma$ has a non-null intersection with $\{\E(Z_x|\cF_t)>0\}$. By \cite[Lemma 2.1.3]{KS}, we deduce an at most  countable subfamily $(x_i)_{i\ge 1}$ such that
the union $\bigcup_i\{\E(Z_{x_i}|\cF_t)>0\}$ is of full measure. Therefore, $Z=\sum_{i=1}^{\infty}2^{-i}Z_{x_i}\ge 0$ is such that $\E(Z|\cF_t)>0$ and we define $Q\ll P$ such that $dQ=(Z/\E(Z|\cF_t))dP$. As the subset $\{ \sum_{u=t+1}^T \theta_{u-1}\Delta S_u,~\theta_{u-1}\in L(\R,\cF_{u-1}) \}$ is a linear vector space contained in $\mathcal{R}_t^T$, we deduce that $(S_u)_{u\in \{t,\ldots, T\}}$ is a $Q$-martingale.

We now prove that 2. implies 3.
Suppose that for every $t\in \{0,\ldots, T\}$, there exists $Q\ll P$ such that $(S_u)_{u \in\{t,\ldots, T\}}$ is a $Q$-martingale with  $\E(dQ/dP|\cF_t)=1$. Let us define, for $u\in \{t,\ldots, T\}$, $\rho_u=\E_{\bP}(dQ/d\bP|\cF_u)$. Then, $\rho_u \geq 0$ and $\rho_t=1$. Consider $\gamma_t\in \mathcal{R}_t^T \cap L^0(\R_+,\cF_t)$, i.e., $\gamma_t$ is $\cF_t$-measurable and is of the form $\gamma_t=\sum_{u=t}^{T-1} \theta_{u}\Delta S_{u+1}-\epsilon_T^+$.
Since $\theta_u$ is $\cF_u$-measurable, $\theta_{u}\Delta S_{u+1}$ admits a generalized conditional expectation under $Q$, knowing $\cF_u$, and we have, by assumption, that $\E_Q(\theta_{u}\Delta S_{u+1}|\cF_u)=0$. The tower law implies that a.s. \vskip -0.8cm

$$\gamma_t=\E_Q(\gamma_t|\cF_t)=\sum_{u=t}^{T-1} \E_Q(\E_Q(\theta_{u}\Delta S_{u+1}|\cF_u)|\cF_t)-\E_Q(\epsilon_T^+|\cF_t)=-\E_Q(\epsilon_T^+|\cF_t).$$\vskip -0.5cm

Hence $\gamma_t=0$ a.s., i.e., AIP holds. Let us show that $\overline{\mathcal{R}_t^T} \cap L^0(\R,\cF_t)\subseteq \mathcal{R}_t^T \cap L^0(\R,\cF_t)$.Consider first a one step model, where $(S_u)_{u\in \{T-1, T\}}$ is a $Q$-martingale with  $\rho_T\geq 0$ and $\rho_{T-1}=1$.
Suppose that $\gamma^n=\theta_{T-1}^n\Delta S_T-\epsilon_T^{n+}\in L^0(\R,\cF_{T})$ converges in probability  to $\gamma^{\infty}\in L^0(\R,\cF_{T-1})$. We need to show that $\gamma^{\infty}\in \mathcal{R}_{T-1}^T \cap L^0(\R,\cF_{T-1})$.

On the $\cF_{T-1}$-measurable set $\Lambda_{T-1}:=\{\liminf_n|\theta_{T-1}^n|<\infty\}$, by \cite[Lemma 2.1.2]{KS}, we may assume w.l.o.g. that $\theta_{T-1}^n$ is convergent to some $\theta_{T-1}^{\infty}$ hence $\epsilon_T^{n+}$ is also convergent and we can  conclude that $\gamma^{\infty}1_{\Lambda_{T-1}}\in \mathcal{R}_{T-1}^T \cap L^0(\R,\cF_{T-1})$. 
Otherwise, on $\Omega\setminus \Lambda_{T-1}$, we use the normalized sequences for $i \in \{1,\ldots,d\}$
$$\tilde \theta_{T-1}^{n,i}:=\theta_{T-1}^{n,i}/(|\theta_{T-1}^n|+1),  \;\; \tilde\epsilon_T^{n+}:=\epsilon_T^{n+}/(|\theta_{T-1}^n|+1).$$ By \cite[Lemma 2.1.2]{KS} again, we may assume, taking $d+1$ sub-sequences, that a.s. $\tilde \theta_{T-1}^n\to \tilde \theta_{T-1}^{\infty}$, $\tilde \epsilon_T^{n+}\to \tilde \epsilon_T^{\infty+}$ and $\tilde \theta_{T-1}^{\infty}\Delta S_T-\tilde \epsilon_T^{\infty+}=0$ a.s. Remark that $|\tilde \theta_{T-1}^{\infty}|=1$ a.s. First consider the subset $\Lambda_{T-1}^2:=\left(\Omega\setminus \Lambda_{T-1}\right)\cap \{\tilde \theta_{T-1}^{\infty}=1\} \in \cF_{T-1}$ on which $\Delta S_T\ge 0$ a.s. Since  $\E_{Q}(\Delta S_T1_{\Lambda_{T-1}^2}|\cF_{T-1})=0$ a.s., we get
that
$\rho_T \Delta S_T1_{\Lambda_{T-1}^2}=0$ a.s. Hence  $\rho_T \gamma^n1_{\Lambda_{T-1}^2}=-\rho_T\epsilon_T^{n+}1_{\Lambda_{T-1}^2} \leq 0$ a.s. Taking the limit, we get that $\rho_T \gamma^{\infty}1_{\Lambda_{T-1}^2}\le 0$ a.s. and, since $\gamma^{\infty}\in L^0(\R,\cF_{T-1})$, we deduce that $\rho_{T-1}\gamma^{\infty}1_{\Lambda_{T-1}^2}\le 0$ a.s. Recall that  $\rho_{T-1}=1$ hence $\gamma^{\infty}1_{\Lambda_{T-1}^2}\le 0$ a.s. and $\gamma^{\infty}1_{\Lambda_{T-1}^2}\in \mathcal{R}_{T-1}^T \cap L^0(\R,\cF_{T-1})$. On the subset $\left(\Omega\setminus \Lambda_{T-1}\right)\cap \{\tilde \theta_{T-1}^{\infty}=-1\}$ we may argue similarly and the conclusion follows in the one step model.

We now show the result in multi-step models by recursion. Fix some $s \in \{t,\ldots,T-1\}$. We show that $\overline{\mathcal{R}_{s+1}^T} \cap L^0(\R,\cF_{s+1})\subseteq \mathcal{R}_{s+1}^T \cap L^0(\R,\cF_{s+1})$ implies the same property for $s$ instead of $s+1$. By assumption $(S_u)_{u\in \{s,\ldots, T\}}$ is a $Q$-martingale with  $\E_{P}(dQ/dP|\cF_u)=\rho_u \geq 0$ for  $u\in \{s,\ldots, T\}$ and $\rho_s=1$. Suppose that
$\gamma^n=\sum_{u=s+1}^{T} \theta_{u-1}^n\Delta S_{u}-\epsilon_T^{n+} \in \mathcal{R}_{s}^T \cap L^0(\R,\cF_T)$ converges to  $\gamma^{\infty}\in L^0(\R,\cF_{s}).$ 
If $\gamma^{\infty}=0$ there is nothing to prove.
As before on the $\cF_{s}$-measurable set $\Lambda_{s}:=\{\liminf_n|\theta_{s}^n|<\infty\}$,  we may assume w.l.o.g. that $\theta_{s}^n$  converges to $\theta_{s}^{\infty}$. Therefore on $\Lambda_{s}$
$$\sum_{u=s+2}^{T} \theta_{u-1}^n\Delta S_{u}-\epsilon_T^{n+}
=\gamma^n - \theta_{s}^n\Delta S_{s+1} \to
\gamma^{\infty} - \theta_{s}^{\infty} \Delta S_{s+1}$$
and, by the induction hypothesis,  $\sum_{u=s+2}^{T} \theta_{u-1}^n\Delta S_{u}-\epsilon_T^{n+}$ also converges to an element of $\mathcal{R}_{s+1}^T\cap L^0(\R,\cF_{s+1})$ and we  conclude that $\gamma^{\infty} 1_{\Lambda_{s}} \in  \mathcal{R}_{s}^T \cap  L^0(\R,\cF_{s})$.\\
On $\Omega\setminus \Lambda_{s-1}$, we use the normalisation procedure as before, and deduce the equality 
$\sum_{u=s+1}^{T} \tilde \theta_{u-1}^{\infty}\Delta S_{u}-\tilde \epsilon_T^{{\infty}+}=0 \mbox{ a.s.}$
for some $\tilde \theta_{u}^{\infty}\in L^0(\R,\cF_{u})$, $u \in \{s,\ldots,T-1\}$ and $\tilde \epsilon_T^{{\infty}+}\ge 0$ such that $|\tilde \theta_{s}^{\infty}|=1$ a.s. We then argue on $\Lambda_{s}^2:=\left(\Omega\setminus \Lambda_{s-1}\right)\cap \{\tilde \theta_{s}^{\infty}=1\}\in \cF_{s}$ and $\Lambda_{s}^3:=\left(\Omega\setminus \Lambda_{s-1}\right)\cap \{\tilde \theta_{s}^{\infty}=-1\}\in \cF_{s}$ respectively.
When $\tilde \theta_{s}^{\infty}=1$, we deduce that
$$\Delta S_{s+1}+\sum_{u=s+2}^{T} \tilde \theta_{u-1}^{\infty}\Delta S_{u} -\tilde \epsilon_T^{{\infty}+}=0 \mbox{ a.s., i.e., } \Delta S_{s+1}\in \cP_{s+1,T}(0)$$ hence $\Delta S_{s+1}\ge \pi_{s+1,T}(0)=0$ a.s. under AIP, see Proposition \ref{thoNip}. Since\\  $\E_{Q}(\Delta S_{s+1}1_{\Lambda_{s}^2}|\cF_{s})=0$ a.s.,   $\rho_{s+1}\Delta S_{s+1}1_{\Lambda_{s}^2}=0$ a.s.
So,
$$\rho_{s+1}\gamma^n 1_{\Lambda_{s}^2}=\sum_{u=s+2}^{T} \theta_{u-1}^n \rho_{s+1} 1_{\Lambda_{s}^2}\Delta S_{u}-\epsilon_T^{n+} \rho_{s+1} 1_{\Lambda_{s}^2}\in \mathcal{R}_{s+1}^T \cap L^0(\R,\cF_{s+1}).$$
Hence, $\rho_{s+1}\gamma^{\infty} 1_{\Lambda_{s}^2} \in \mathcal{R}_{s+1}^T \cap L^0(\R,\cF_{s+1})$ by induction.
As $\rho_{s+1}\gamma^{\infty}1_{\Lambda_{s}^2}$ admits a generalized conditional expectation knowing $\cF_s$, the tower property implies  a.s.
\begin{eqnarray*}
1_{\Lambda_{s}^2} \E(\rho_{s}\gamma^{\infty}|\cF_s) & = &
\E(\rho_{s+1}\gamma^{\infty}1_{\Lambda_{s}^2}|\cF_s) 
 = 
\sum_{u=s+2}^{T} 1_{\Lambda_{s}^2} \E\left(\theta_{u-1}^{\infty}\E  \left(\frac{dQ}{dP}\Delta S_{u}|\cF_{u-1}\right) |\cF_s\right) \\
 & & -1_{\Lambda_{s}^2} \E(\epsilon_T^{\infty+} \rho_{s+1} |\cF_s) 
 \leq  0,
\end{eqnarray*}
since $(S_u)_{u\in \{s,\ldots, T\}}$ is a $Q$-martingale. Hence, $\rho_{s}\gamma^{\infty}1_{\Lambda_{s}^2}\le 0$ a.s.  As  $\rho_{s}=1,$ $\gamma^{\infty}1_{\Lambda_{s}^2}\le 0$ a.s. so that $\gamma^{\infty}1_{\Lambda_{s}^2}\in \mathcal{R}_{s}^T \cap L^0(\R,\cF_{s})$.

 Finally, notice that the AIP condition implies AWIP  as soon as the equality $\overline{\mathcal{R}_t^T} \cap L^0(\R_+,\cF_t)=\mathcal{R}_t^T \cap L^0(\R_+,\cF_t)$ holds for every $t \in \{0,\ldots,T\}$.

Suppose now that $P(\essinf[\cF_t]S_{t+1}=S_t)=P(\esssup[\cF_t]S_{t+1}=S_t)=0$. Then, using Proposition \ref{NAAIP},
 AIP is equivalent to NA. Under NA, the set $\mathcal{R}_{t}^T$ is closed in probability for every $t\in \{0\ldots,T-1\}$ and
 what we have just proved implies that AWIP, AIP and NA are equivalent conditions.
 \fdem \\

\section*{Additional informations}
A former version of the paper was called "Pricing without martingale measure" and was co-authored with Julien Baptiste. The paper has been split into two parts. This paper  contains the theoretical study of the superreplication problem while a second one, with Julien Baptiste, focuses on numerical aspects (see, \cite{BCL}).

\end{document}